\documentclass[acmsmall,authorversion,screen]{acmart} 

\AtBeginDocument{%
  }

\setcopyright{cc}
\setcctype{by}
\acmDOI{10.1145/3729302}
\acmYear{2025}
\acmJournal{PACMPL}
\acmVolume{9}
\acmNumber{PLDI}
\acmArticle{199}
\acmMonth{6}
\received{2024-11-14}
\received[accepted]{2025-03-06}

\usepackage{prelude}


\begin{document}

\title{An Interactive Debugger for Rust Trait Errors}

\author{Gavin Gray}
\orcid{0000-0002-2960-1198}
\email{gavin_gray@brown.edu}

\author{Will Crichton}
\orcid{0000-0001-8639-6541}
\email{will_crichton@brown.edu}

\author{Shriram Krishnamurthi}
\orcid{0000-0001-5184-1975}
\affiliation{%
  \institution{Brown University}
  \city{Providence}
  \country{USA}
}

\setlength{\enumerateparindent}{\parindent}

\begin{abstract}
Compiler diagnostics for type inference failures are notoriously bad, and type classes only make the problem worse. By introducing a complex search process during inference, type classes can lead to wholly inscrutable or useless errors. We describe a system, \argus, for interactively visualizing type class inferences to help programmers debug inference failures, applied specifically to Rust's trait system.  The core insight of \argus is to avoid the traditional model of compiler diagnostics as one-size-fits-all, instead providing the programmer with different views on the search tree corresponding to different debugging goals. \argus carefully uses defaults to improve debugging productivity, including interface design (e.g., not showing full paths of types by default) and heuristics (e.g., sorting obligations based on the expected complexity of fixing them). We evaluated \argus in a user study where \estimate{$N = 25$} participants debugged type inference failures in realistic Rust programs, finding that participants using \argus correctly localized \estimate{\LocalizedRateFactor} as many faults and localized \estimate{\LocalizedTimeFactor} faster compared to not using \argus.
\end{abstract}

  
\begin{teaserfigure}
  \includegraphics[width=\textwidth]{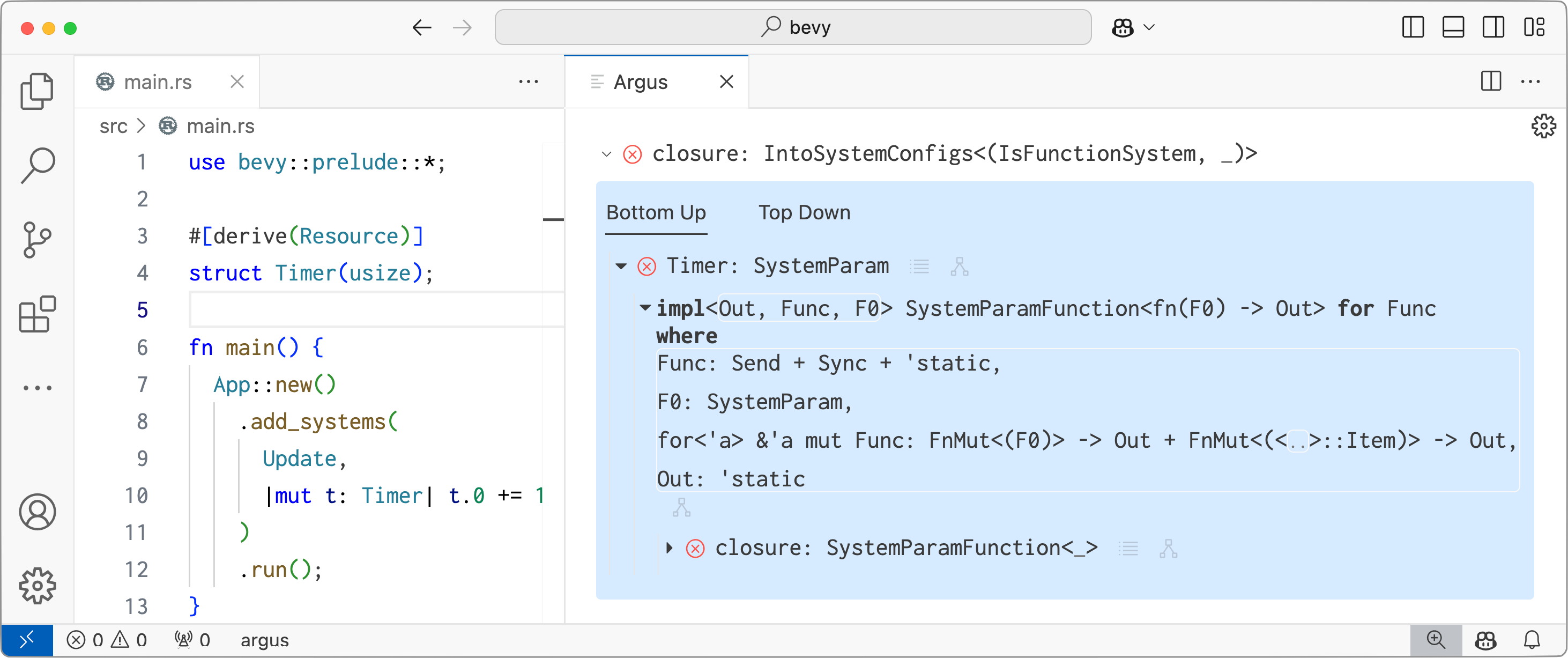}
  \caption{A screenshot of the \argus{} trait debugger's bottom-up view in VS Code applied to a Bevy program (\Cref{sec:case-study-bevy}). Using an interactive graphical interface for the trait inference tree, \argus{} can include key information (the bound \tcbox[verbatim,colback=white,colframe=codeborder,boxsep=0mm,left=2pt,right=2pt,top=2pt,bottom=2pt]{\lstinline^Timer: SystemParam^}) elided by the Rust compiler diagnostic for the same program.}
  \label{fig:teaser}
\end{teaserfigure}

\maketitle

\section{Introduction}

Type classes improve the brevity of bounded polymorphism by implicitly passing inferred type class instances to generic functions, as opposed to ML-style explicit passing of modules via functors. In other words, type classes shift the burden of finding instances from the programmer to the compiler. When a type class inference succeeds, the programmer does not need to expend any thought on the inference process, at least for systems which enforce coherence~\cite{bottu2019coherence,reynolds1991coherence}. But when a type class inference fails, the compiler is responsible for explaining the failure to the programmer.

Back in the days when type classes were a simple tool for overloading~\cite{wadler1989adhoc}, diagnostics posed no particular challenge. Say a function has a type like \rs{ToString a => a -> String}, and you pass an \rs{int}, and there happens to be no instance of \rs{ToString} for \rs{int}. Then the compiler just needs to say: ``no instance found for \rs{ToString int}''.

The challenge today is that type class systems are more powerful than before. Type classes in Haskell, Coq, Rust (traits), and Scala (implicits) have all been shown to encode Turing-complete computations. Rust even explicitly models type class inference (``trait solving'') as Prolog-esque logic programming~\cite{chalk}. Library developers in these languages, especially Rust, increasingly lean on type classes to encode domain-specific correctness properties into the type system. While this approach helps developers catch more mistakes at compile-time, it can at times produce mystifying diagnostics. Given the severity of this problem, the Rust community has invested money into studying trait diagnostics~\cite{weiznich-grant} as well as developed both library-specific~\cite{bevycheck} and library-agnostic~\cite{diagnostic-rfc} utilities solely for improving trait diagnostics.

The thesis of this work is that compiler diagnostics are fundamentally limited by their representation as static text. Moreover, this limitation is felt most acutely for information-rich situations such as type class inference. We therefore designed a system, \argus, to provide a richer interface for explaining type class inferences built on a modern UI framework. \argus is implemented as an IDE extension for Rust, although its core design is not particularly Rust-specific. After motivating our design principles with concrete examples (\Cref{sec:case-studies}), we describe our contributions:
\begin{itemize}
    \item A novel interface for visualizing trait inference, designed to specifically facilitate key sub-tasks in debugging inference failures (\Cref{sec:interface-design}).

    \item A new heuristic, \textit{inertia}, that ranks potential root causes of trait inference failure (\Cref{sec:inertia}).  

    \item A user study that shows that participants using \argus could localize faults \estimate{\LocalizedTimeFactor} faster compared to using the Rust compiler's diagnostics (\Cref{sec:evaluation}).
\end{itemize}

\section{Motivating Examples}\label{sec:case-studies}

Traits are a well-documented source of confusing compiler errors in the Rust community. A 2023 study~\cite{weiznich-grant} commissioned by the Rust Foundation identified dozens of problematic error messages in widely-used libraries, resulting in a corpus of hard-to-debug programs. We analyzed the content of these error messages to form hypotheses about how to design better trait diagnostics. 

We chose three programs that represent the main failure modes of programs in this corpus. For illustration, the program in \Cref{sec:case-study-ast} was taken from an online Rust forum~\cite{discourse-question} because it requires less boilerplate than programs in the corpus, but nonetheless contains an equivalent error. We start by walking through these concrete examples that illustrate the problems in Rust's existing trait diagnostics. We then generalize these examples into design principles that form the basis of the \argus interface. 
All errors in this section were generated using \estimate{Rust 1.82.0}.

\subsection{A Missing Table Join}
\label{sec:case-study-diesel}
\begin{figure}
    \centering
    \begin{subfigure}{\linewidth}
    \centering
\begin{minipage}{0.8\linewidth}
\begin{lstlisting}[aboveskip=0pt,belowskip=0pt]
fn users_with_eq_post_id(conn: &mut PgConnection) -> Vec<(i32, String>> {
  users::table // .inner_join(posts::table)
    .filter(users::id.eq(posts::id))
    .select((users::id, posts::id))
    .load::<(i32, String)>(conn)
}
\end{lstlisting} 
\end{minipage}
    \cprotect\caption{A program using the Diesel query builder library. The program does not join the table \rs{posts} but tries to use the \rs{posts::id} column, which Diesel catches as a trait error.}
    \label{fig:diesel-backtrace-code}
    \end{subfigure}
    \begin{subfigure}{\linewidth}
\begin{lstlisting}[basicstyle=\ttfamily\scriptsize,language={},numbers=none,belowskip=0pt]
error[E0271]: type mismatch resolving `<table as AppearsInFromClause<table>>::Count == Once`
     |
 5   |         .load::<(i32, String)>(conn);
     |          ----                  ^^^^ expected `Once`, found `Never`
     |          |
     |          required by a bound introduced by this call
     |
note: required for `posts::columns::id` to implement `AppearsOnTable<users::table>`
     |
18   |         id -> Integer,
     |         ^^
     = note: associated types for the current `impl` cannot be restricted in `where` clauses
     = (*@\highlight{note: 2 redundant requirements hidden}@*)
     = note: required for `diesel::expression::grouped::Grouped<diesel::expression::operators::Eq<users::columns::id, posts::columns::id>>` to implement `AppearsOnTable<users::table>`
     = note: required for `query_builder::where_clause::WhereClause<diesel::expression::grouped::Grouped<diesel::expression::operators::Eq<users::columns::id, posts::columns::id>>>` to implement `query_builder::where_clause::ValidWhereClause<FromClause<users::table>>`
     = note: required for `SelectStatement<FromClause<table>, SelectClause<(id, name)>, NoDistinctClause, WhereClause<Grouped<Eq<id, id>>>>` to implement `Query`
     = note: required for `SelectStatement<FromClause<table>, SelectClause<(id, name)>, NoDistinctClause, WhereClause<Grouped<Eq<id, id>>>>` to implement `LoadQuery<'_, _, (i32, String)>`
note: required by a bound in `diesel::RunQueryDsl::load`
     |
1540 |     fn load<'query, U>(self, conn: &mut Conn) -> QueryResult<Vec<U>>
     |        ---- required by a bound in this associated function
1541 |     where
1542 |         Self: LoadQuery<'query, Conn, U>,
     |               ^^^^^^^^^^^^^^^^^^^^^^^^^^ required by this bound in `RunQueryDsl::load`
\end{lstlisting}
    \caption{The Rust compiler diagnostic for the program above.}
    \label{fig:diesel-backtrace-diagnostic}
    \end{subfigure}
    \caption{An example of a trait error where the compiler elides key information for brevity. This is noted by the phrase ``2 redundant requirements hidden.''}
    \label{fig:diesel-backtrace}
    \vspace{-1em}
\end{figure}
\noindent Diesel~\cite{diesel} is a popular Rust library for object-relational mapping and statically-checked query building. \Cref{fig:diesel-backtrace-code} shows an example Diesel program where a developer wants to select fields from two tables, \rs{users} and \rs{posts}, but forgot to join the \rs{posts} table into the query. Diesel uses traits to discover that the call to \rs{.load(conn)} is ill-typed because the query selects a field of a missing table, generating the diagnostic shown in \Cref{fig:diesel-backtrace-diagnostic}.
To debug this trait error, a developer must \emph{localize} the root cause (i.e., the \rs{.eq(post::id)} operation) and \emph{fix} the program (e.g., by inserting a join).

The diagnostic's goal is principally to help with the localization phase of debugging by providing context about the origin of the type error. For instance, the Rust compiler diagnostic in \Cref{fig:diesel-backtrace-diagnostic} does not just report the top-level failed trait bound, printed at the very bottom of the diagnostic. Rust instead starts by reporting the failed predicate deepest into the trait inference tree:
\begin{lstlisting}[language={},numbers=none]
type mismatch resolving `<table as AppearsInFromClause<table>>::Count == Once`
\end{lstlisting}

The developer's localization task is to blame a specific program element for this failed predicate. This task presents two problems. First, the associated type \rs{AppearsInFromClause::Count} may not be self-evidently meaningful, requiring additional context to interpret the constraint (e.g., where did this constraint come from?). Second, the types \rs{users::table} and \rs{posts::table} have been unfortunately truncated to simply \rs{table}, suggesting the types are the same when they are not.

To try solving problem \#1, the developer could read the rest of the diagnostic. The remaining text explains the provenance of the constraint, which is a sequence of five trait constraints deriving from the originating constraint on the call to \rs{.load(conn)}. One possibly useful constraint to read would be \rs{Eq<users::columns::id, posts::columns::id>: AppearsOnTable<users::table>}.
The \rs{Eq<...>} type hints at the problem originating with the expression \rs{users::id.eq(posts::id)}. This bound helps the developer localize the fault: that \rs{posts} must be joined before using this expression. 

However, this constraint does not actually appear in the text of the diagnostic! It is elided with the statement: \rs{= note: 2 redundant requirements hidden}.
Also observe that the diagnostic includes the source code and location for the first two trait bounds walking up from the deepest failed bound (at the top of the diagnostic) and the originating trait bound (at the bottom of the diagnostic). The diagnostic does not include this information for any of the intermediate trait bounds.

The Rust compiler omits all this information out of necessity, not convenience. Consider the counterfactual where Rust includes the full text of every bound and its source-mapped origin. This diagnostic could easily stretch over 100 lines long, just for a relatively simple error. Therefore, Rust applies heuristics to include only information that is probably relevant. The problem with identical-looking \rs{table} types is similar. Rust heuristically decides when to present fully-qualified versus shortened paths for brevity, but it sometimes makes a wrong decision.
Without representing diagnostics as static text, we can consider alternative solutions to both problems:

\principle{1}{\prinzipEis}{Instead of omitting steps of an inference sequence for brevity, allow the developer to progressively unfold the sequence.}

\principle{2}{\prinzipZwoo}{Instead of heuristically shortening types, show shortened types by default, but make fully-qualified types available on-demand.}

\begin{figure}
    \begin{minipage}{0.4\linewidth}
    \begin{subfigure}{\linewidth}
\begin{lstlisting}[aboveskip=0pt,belowskip=0pt]
trait AssocData<A: AstAssocs> {}
trait AstAssocs: Sized {
  type Data: AssocData<Self>;
}

struct EmptyNode;
struct Statement<A: AstAssocs>(..);

impl<Data> AstAssocs for Data
where Data: AssocData<Self> {
  type Data = Data;
}

impl<A> AssocData<A> for EmptyNode 
where A: AstAssocs {}

fn main() {
  let s: Statement<EmptyNode> = 
    Statement(..);
}
\end{lstlisting}
        \caption{A program which tries to model an AST with user-specified associated data on AST nodes. The described trait bounds and impl blocks cause an infinite loop in the trait solver.}
        \label{fig:ast-overflow-code}
    \end{subfigure}
    \end{minipage}
    \hfill
    \begin{minipage}{0.59\linewidth}
        \begin{subfigure}{\linewidth}
            \begin{lstlisting}[numbers=none,language={},aboveskip=0pt,belowskip=0pt]
error[E0275]: overflow evaluating the requirement `EmptyNode: AssocData<EmptyNode>`
   |
18 |   let s: Statement<EmptyNode> =
   |          ^^^^^^^^^^^^^^^^^^^^
   |
note: required for `EmptyNode` to implement `AstAssocs`
   |
 9 | impl<Data> AstAssocs for Data
   |            ^^^^^^^^^     ^^^^
10 | where Data: AssocData<Self> {
   |             --------------- 
       unsatisfied trait bound introduced here
note: required by a bound in `Statement`
   |
 7 | struct Statement<A: AstAssocs>(
   |                     ^^^^^^^^^ 
          required by this bound in `Statement`
\end{lstlisting}
            \caption{The Rust compiler diagnostic for the program on the left.}
            \label{fig:ast-overflow-diagnostic}
        \end{subfigure}
    \begin{subfigure}{\linewidth}
            \begin{align*}
                    &\mtt{EmptyNode: AstAssocs} \\[-0.3em]
                    \overset{\mtt{impl 9-12}}{\impliedby} &\mtt{EmptyNode: AssocData<EmptyNode>} \\[-0.3em]
                    \overset{\mtt{impl 14-15}}{\impliedby} &\mtt{EmptyNode: AstAssocs}
            \end{align*}            
            \caption{A diagrammatic representation of the logical structure of the recursion.}
            \label{fig:ast-overflow-diagram}
        \end{subfigure}
    \end{minipage}    
    \caption{An example of a trait error where the interleaving of information in the diagnostic obscures the logical structure of the problem.}
    \label{fig:ast-overflow}
    \vspace{-1em}
\end{figure}

\subsection{An Accidental Infinite Recursion}
\label{sec:case-study-ast}

A Rust developer was designing an AST data type to be generic with respect to user-specific node-associated data. They wrote the code in \Cref{fig:ast-overflow-code}, which caused an infinite loop in the trait solver, as indicated in the trait diagnostic in \Cref{fig:ast-overflow-diagnostic}. The developer asked on a Rust forum~\cite{discourse-question}:
\begin{quote}
    I'm running into a compiler error, stating that there is an overflow when evaluating a trait requirement. However, that requirement should obviously be satisfied. I just can't seem to understand where the overflow comes from.
\end{quote} 

The actual loop has a simple logical structure, as shown in \Cref{fig:ast-overflow-diagram}. If  \rs{EmptyNode} needs to implement \rs{AstAssocs} (due to line 18), then that requires \rs{EmptyNode} implements \rs{AssocData<EmptyNode>} (due to line 10), which in turn requires \rs{EmptyNode} implements \rs{AstAssocs} (due to line 15). 

However, the Rust diagnostic obscures this fact because the diagnostic interleaves the ``core'' information used in the trait solver (the trait-bounds and impl blocks) with ``auxiliary'' information used for debugging (the source-location of constraints). This approach makes it harder for a developer to identify the logical structure of the cycle. Again, the ultimate issue is the static text representation, which requires diagnostics to commit to a specific sequential interleaving of all relevant information. Therefore our principle is:

\principle{3}{\prinzipDruu}{Instead of interleaving the trait inference steps with auxiliary information, enable developers to access auxiliary information on-demand through contextual links.}

\begin{figure}
    \centering
    \begin{subfigure}[t]{0.43\textwidth}
\begin{lstlisting}[aboveskip=0pt,belowskip=0pt]
#[derive(Resource)]
struct Timer(usize);

fn run_timer(
//mut timer: ResMut<Timer>
  mut timer: Timer
) { timer.0 += 1; }

fn main() {
  App::new()
    .insert_resource(Timer(0))
    .add_systems(Update, run_timer)
    .run();
}
\end{lstlisting}
        \cprotect\caption{A program using the Bevy game engine. The \rs{run_timer} function incorrectly takes a parameter of type \rs{Timer} instead of \rs{ResMut<Timer>}.}
        \label{fig:bevy-example-code}
    \end{subfigure}
    \hfill
    \begin{subfigure}[t]{0.55\linewidth}
\begin{lstlisting}[aboveskip=0pt,language={},numbers=none,belowskip=0pt]
error[E0277]: `fn(Timer) {run_timer}` does not describe a valid system configuration
    |
12  |   .add_systems(Update, run_timer)
    |    -----------         ^^^^^^^^^ 
    |    |          invalid system configuration
    |    |
    |  required by a bound introduced by this call
    |
= help: the trait `IntoSystem<(), (), _>` is not implemented for fn item `fn(Timer) {run_timer}`, which is required by `fn(Timer) {run_timer}: IntoSystemConfigs<_>`
\end{lstlisting}        
    \caption{The Rust compiler diagnostic for the program on the left.}
    \label{fig:bevy-example-diagnostic}
    \end{subfigure}
    \begin{subfigure}{\textwidth}
        \centering
        \includegraphics[width=\textwidth]{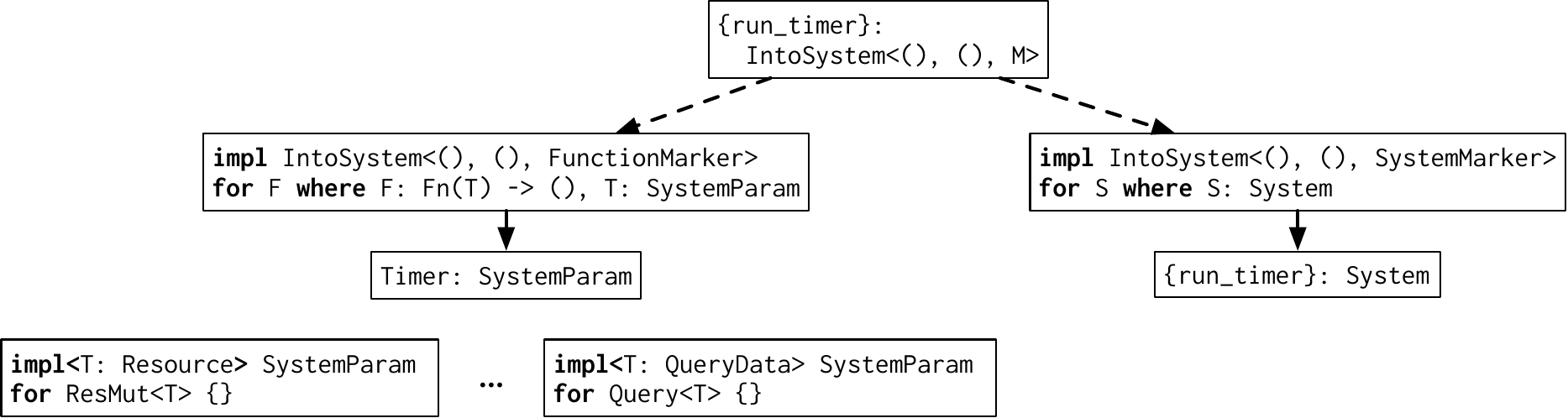}
        \cprotect\caption{A diagrammatic representation of the relevant fragment of the trait inference tree, showing the branch point in the possible implementations of the \rs{IntoSystem} trait.}
        \label{fig:bevy-example-diagram}
    \end{subfigure}
    \caption{An example of a trait error where a branch point in the inference process causes the diagnostic to omit information deeper in the search tree.}
    \label{fig:bevy-vignette}
\end{figure}

\subsection{An Errant Function Parameter}\label{sec:case-study-bevy}

Bevy~\cite{bevy} is a popular Rust library for writing 2D and 3D games in the entity-component-system (ECS) style. Systems in ECS are functions that perform updates on the game. A system's function parameters declare the required inputs to the system, and the game engine essentially does dependency injection to run the system with the appropriate inputs.

For example, \Cref{fig:bevy-example-code} shows a developer trying to write a system which increments a global mutable timer. The correct approach is to declare \rs{Timer} as a \rs{Resource}, and then to use the container type \rs{ResMut<Timer>} as the function parameter. However, a common Bevy mistake is to forget the container and simply write \rs{timer: Timer}. The function \rs{run_timer} is still well-typed, but now Bevy rejects the developer's attempt to register the system on the game.
The type error arises from a failed trait inference, where the method \rs{add_systems} requires that \rs{run_timer} implements a trait \rs{IntoSystem}, which converts a value into a system. A function system requires each parameter to implement a trait \rs{SystemParam}, for which one implementation is that \rs{ResMut<T>: SystemParam} if \rs{T: Resource}. \rs{Timer} does not implement \rs{SystemParam}, so \rs{run_timer} does not implement \rs{IntoSystem}.

The problem is that the Rust diagnostic, shown in \Cref{fig:bevy-example-diagnostic}, only mentions the \rs{IntoSystem} bound. It says, essentially, ``something is wrong with the type of \rs{run_timer}'' without pointing to a more specific culprit. Rust lacks specificity because there are other ways to potentially implement \rs{IntoSystem} for \rs{run_timer}. The diagram in \Cref{fig:bevy-example-diagram} shows how \rs{IntoSystem} can also be implemented for types that implement the \rs{System} trait.\footnotemark

\cprotect\footnotetext{These implementations seems to violate coherence. Indeed, with a straightforward definition of \rss{IntoSystem}, Rust would reject the two impl blocks as overlapping. Bevy employs the technique of adding a \emph{marker type parameter} to the \rss{IntoSystem} trait, written as \rss{FunctionMarker} and \rss{SystemMarker} in \Cref{fig:bevy-example-diagram}. This parameter ensures that the two implementations are not strictly overlapping. It then increases the burden on Rust's type inference to deduce the correct type of the marker.}

Once more, we observe the limitations of the static text representation. Rust adopts the approach that when a branch point exists in the trait inference tree, its diagnostics stop at the branch point and do not provide finer-grained details along every branch. In other words, the diagnostics are constrained to presenting a \emph{sequence} of information, not a \emph{tree} of information. Therefore, we adopt the principle:

\principle{4}{\prinzipVier}{Instead of omitting tree-shaped information in a trait inference, provide an interface that supports exploring trait inference as a tree.}

\section{System Design}
\label{sec:conceptual-model}

\argus{} facilitates trait debugging by visualizing the entire trait inference tree in an interactive interface. The primary goal of \argus{} is to help developers localize the root cause of trait errors, i.e., specific failed trait obligations. \argus{} consists of two principal components: 
\begin{enumerate}
    \item A Rust compiler plugin that extracts an idealized representation of trait inferences.
    \item A web-based interface for visualizing extracted trait inferences inside an IDE.
\end{enumerate}

In this section, we describe the concepts most fundamental to \argus{}: an idealized representation of trait inferences (\Cref{sec:model}), the interface design (\Cref{sec:interface-design}), and the heuristics used to organize information in the interface (\Cref{sec:inertia}). We discuss the implementation details of extracting trait inferences in \Cref{sec:implementation}.

\subsection{Trait Model}\label{sec:model}

\begin{figure}
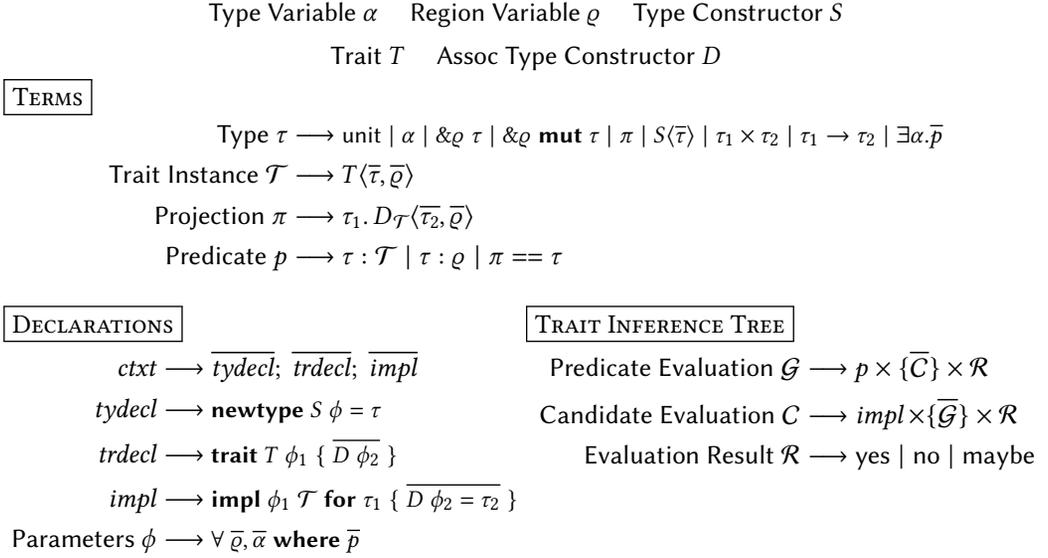

    \centering
    \begin{center}
    $$
        \msf{Type~Variable}~\tvar \hspace{12pt}\msf{Region~Variable}~\lvar\hspace{12pt} \msf{Type~Constructor}~S
    $$
    $$
        \msf{Trait}~T \hspace{12pt} \msf{Assoc~Type~Constructor}~D
    $$
    \end{center}
    \begin{minipage}{\textwidth}
    \synhead{Terms}{0em}
    $$
    \begin{aligned}
    \bnfm{Type}{\term}{\tunit \mid \tvar \mid \&\lvar~\term \mid \&\lvar~\msfb{mut}~\term \mid \pi \mid S\tapp{\seqof{\term}} \mid \term_1\times\term_2 \mid \term_1\rightarrow\term_2 \mid \exists \alpha.\seqof{p}} \\
    \bnfm{Trait~Instance}{\traitinst}{T\tapp{\seqof{\term}, \seqof{\lvar}}} \\
    \bnfm{Projection}{\pi}{\term_1.\,D_\traitinst\tapp{\seqof{\term_2}, \seqof{\lvar}}} \\
    \bnfm{Predicate}{p}{\term: \traitinst \mid \term: \lvar \mid \pi == \term} \\
    \end{aligned}
    $$
    \end{minipage}
    \vspace{\bigskipamount}
    
    \begin{minipage}[t]{0.5\textwidth}
    \synhead{Declarations}{0em} 
    $$
    \begin{aligned}
    \bnfm{}{\ctxt}{\seqof{\tydecl};~\seqof{\trdecl};~\seqof{\impl}} \\
    \bnfm{}{\tydecl}{\msfb{newtype}~S~\params = \term} \\
    \bnfm{}{\trdecl}{\msfb{trait}~T~\params_1~\{~\seqof{D~\params_2}~\}} \\
    \bnfm{}{\impl}{\msfb{impl}~\params_1~\traitinst~\msfb{for}~\term_1~\{~\seqof{D~\params_2 = \term_2}}~\} \\
    \bnfm{Parameters}{\params}{\forall~\seqof{\lvar},\seqof{\tvar}~\msfb{where}~~\seqof{p}} \\
    \end{aligned}
    $$
    \end{minipage}\begin{minipage}[t]{0.5\textwidth}
    \synhead{Trait Inference Tree}{0pt}
    $$
    \begin{aligned}
    \bnfm{Predicate~Evaluation}{\Goal}{p \times \SetOf{\mathcal{C}} \times \Result} \\
    \bnfm{Candidate~Evaluation}{\Candidate}{\impl \times \SetOf{\Goal} \times \Result} \\
    \bnfm{Evaluation~Result}{\Result}{\msf{yes} \mid \msf{no} \mid \msf{maybe}}
    \end{aligned}
    $$
    \end{minipage}

    \caption{A grammar for \tLang, the essence of Rust's trait language and inference.}
    \label{fig:the-essence-of-traits}
\end{figure}

First, we need to describe the precise shape of a trait inference to understand what is being visualized in the \argus{} interface. \argus{} operates over a \emph{trait language}, which is the subset of Rust features relevant to trait inference. This trait language is embedded within a \emph{trait inference tree} extracted from the compiler, which can be conceptualized as a partial proof in a natural deduction system.

\Cref{fig:the-essence-of-traits} describes \tLang, the core syntax of Rust's trait language and trait inference trees. Rust's trait language consists of types $\tau$, which are mostly standard with the notable addition of region-annotated references. Types are embedded in declarations, which include newtypes ($\tydecl$), traits ($\trdecl$), and implementation blocks ($\impl$). Newtypes are relevant to the model because nominal typing permits otherwise overlapping trait implementations for the same type.

At a high level, the semantics of traits are that given a context $\ctxt$ and a predicate $p$, the compiler produces a trait inference tree $\Goal$ which describes either a successful or failed inference. An evaluated predicate $\Goal$ consists of the predicate $p$, a result \Result, and a set of evaluated candidates $\Candidate$. If the predicate definitely succeeded or failed then the result is \yes or \no. If a predicate refers to an un-inferred type variable, then the result is \maybe.
A predicate evaluation succeeds if one of its candidates succeeds, which in turn succeeds if all of its nested predicates succeed. Therefore a trait inference tree is an ``\AndOr tree,'' of the same type found in logic program execution.

It is beyond the scope of this paper to provide a formal semantics for \tLang, e.g., a description of the trait solving process or coherence checks. The core design of \argus{} is largely agnostic to the internal details of the trait solver --- our main focus is how to visualize the inference tree once extracted from the compiler. Instead, we will observe a few key facts about Rust's type class design which influence the kinds of trait inference trees that can emerge:

\begin{itemize}
    \item Rust supports multi-parameter type classes. A trait can be instantiated with type parameters, and each instance is distinct from the others for purposes of coherence.
    \item Rust supports flexible instances and flexible contexts. Any type can be used in the ``head'' or ``self'' of an impl block and the constraints of a trait definition, so long as coherence is satisfied.
    \item Rust supports undecidable instances. It places no restrictions on the kinds of recursion permitted in trait bounds, as shown in \Cref{sec:case-study-ast}.
\end{itemize}

Note that we call \tLang an \emph{idealized} model of Rust's trait language for two reasons. First, the model omits features that are part of Rust's type system but don't meaningfully affect the design of \argus{}, such as constant value generics. Second, the model abstracts the complexity of the trait solver, which does not actually produce the beautiful \AndOr tree shown in \Cref{fig:the-essence-of-traits}. We describe how to bridge that gap in \Cref{sec:implementation}.

\subsection{Interface Design}
\label{sec:interface-design}

The \argus{} interface, shown in \Cref{fig:teaser}, takes an evaluated predicate $\Goal$ and presents an interactive visualization of $\Goal$ to the developer. The \argus{} interface is embedded in an IDE extension (specifically to VS Code, in our prototype) which opens a window adjacent to the developer's code when the developer's program contains a trait error. The \argus{} interface can also be embedded in other contexts, such as in an online textbook to pedagogically illustrate the process of trait inference in the style of recent work~\cite{cgk:ownership-conceptual-model}.

\argus{} is principally inspired by performance profiling tools, which also visualize a different kind of tree: a profile, i.e., a weighted call graph. Profilers can generally visualize a profile in either a top-down way (starting at the main function and descending to callees) or a bottom-up way (starting at the functions deepest in the call graph, and ascending to callers). \argus{} similarly exposes top-down and bottom-up views on the trait inference tree.

The details of the \argus{} interface are inspired by the design principles described in \Cref{sec:case-studies}, which we elaborate below.

\begin{figure}
    \begin{subfigure}[t]{0.45\textwidth}
        \boxf{\includegraphics[width=\textwidth]{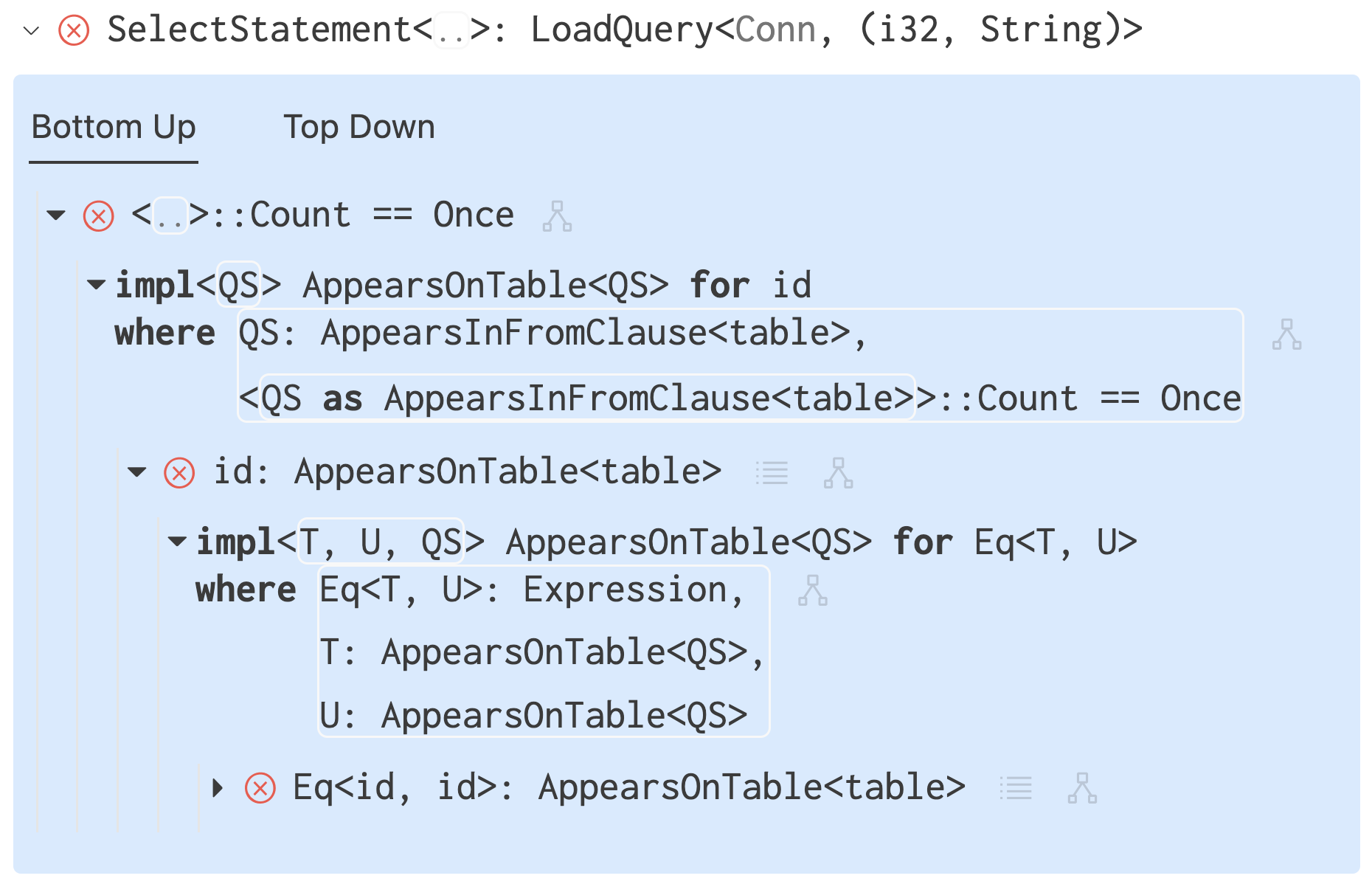}}
        
        \caption{Expanding the inference tree in the bottom-up view.}
        \label{fig:diesel-backtrace-argus}
    \end{subfigure}
    \hfill
    \begin{subfigure}[t]{0.53\textwidth}
        \boxf{\includegraphics[width=\textwidth]{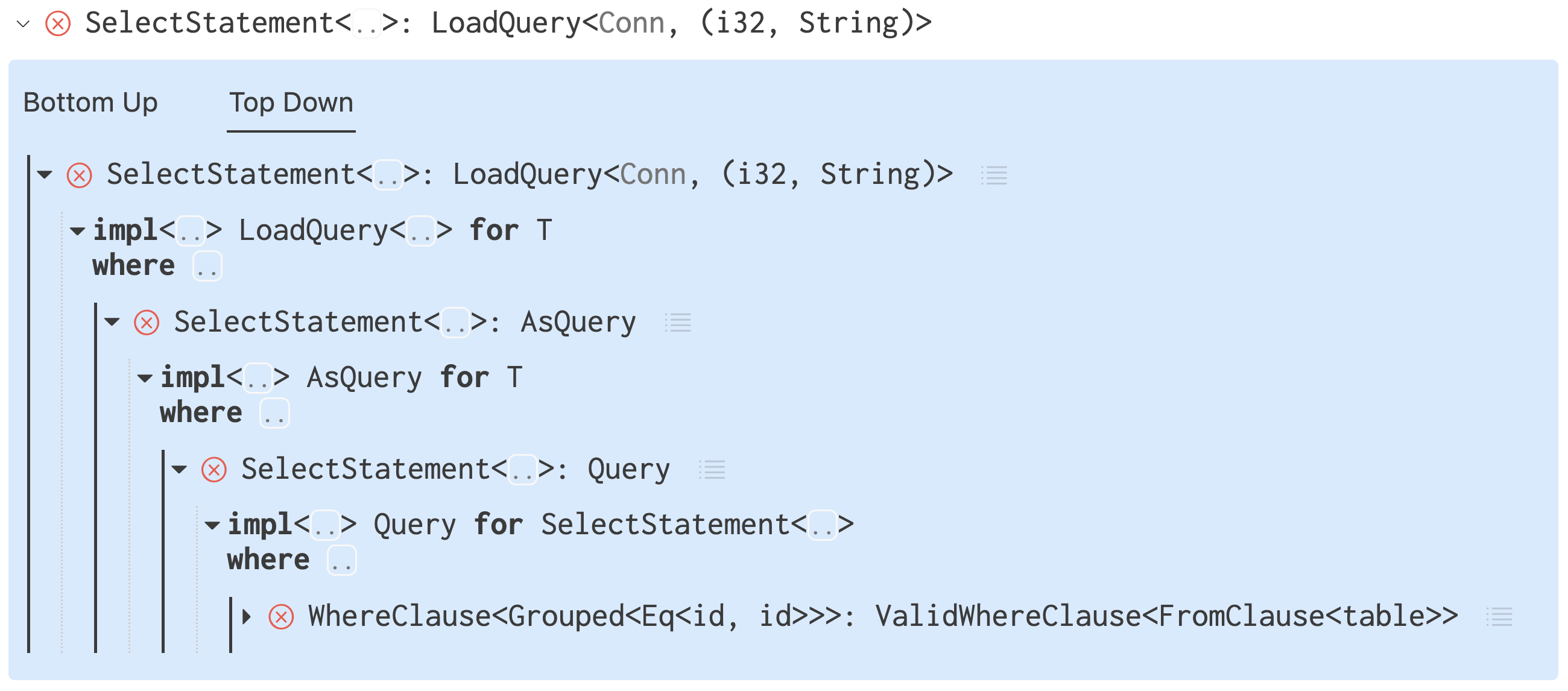}}
        \caption{Expanding the inference tree in the top-down view.}
        \label{fig:diesel-forwardtrace-argus}
    \end{subfigure}
    \vspace{-1em}
    \caption{Interactions in \argus{} for iteratively expanding inference steps.}
\end{figure}

\subsubsection{\prinzipEis}\label{principle:1}
\textit{Instead of omitting steps of an inference sequence for brevity, allow the developer to progressively unfold the sequence.}

Unlike traditional compiler diagnostics, \argus{} presents an \emph{exhaustive} view onto the trait inference tree. Every node is accessible with enough user interaction. To avoid overwhelming the developer with information, the developer iteratively unfolds levels of the tree. \argus provides two views onto the trait inference tree: bottom-up and top-down. 

The bottom-up view shows the leaves of the tree first and developers can recursively expand downward towards the tree root. The developer traverses the tree from the bottom up. For example, \Cref{fig:diesel-backtrace-argus} shows the bottom-up view in \argus{} for the Diesel type error discussed in \Cref{sec:case-study-diesel}. Starting at the innermost failed trait bound, the developer can recursively expand its children (i.e., its parents in the inference tree) until reaching a trait bound that provides useful information about the situation, such as the \rs{Eq<...>} type. 

The top-down view first shows the root of the tree, i.e., the required trait bound in the program, and developers can recursively expand the children until reaching the tree leaves. For example, \Cref{fig:diesel-forwardtrace-argus} shows the top-down view in \argus for the same Diesel type error.

\begin{figure}
    \begin{subfigure}{0.45\textwidth}
        \boxf{\includegraphics[width=\textwidth]{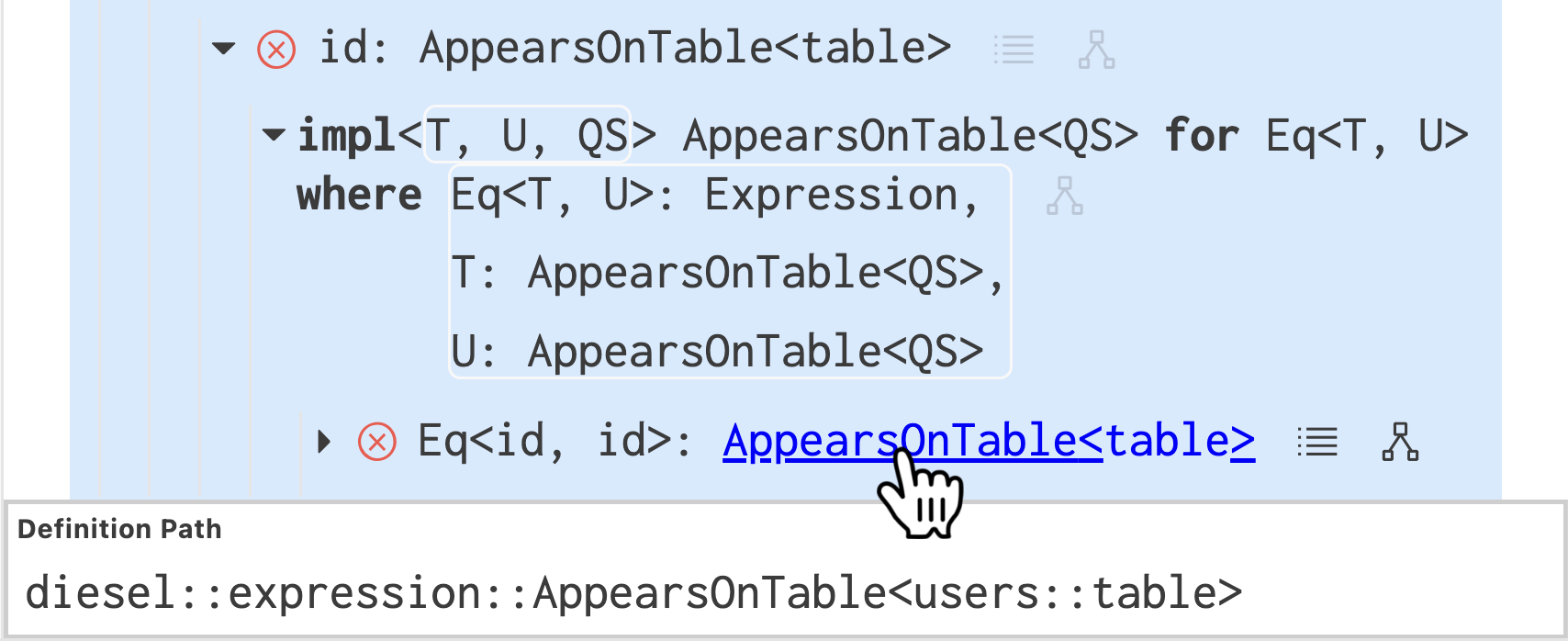}}
        \caption{Hovering over a type to see its fully-qualified paths in the minibuffer.}
        \label{fig:argus-minibuffer}
    \end{subfigure}
    \hfill
    \begin{subfigure}{0.45\textwidth}
        \boxf{\includegraphics[width=\textwidth]{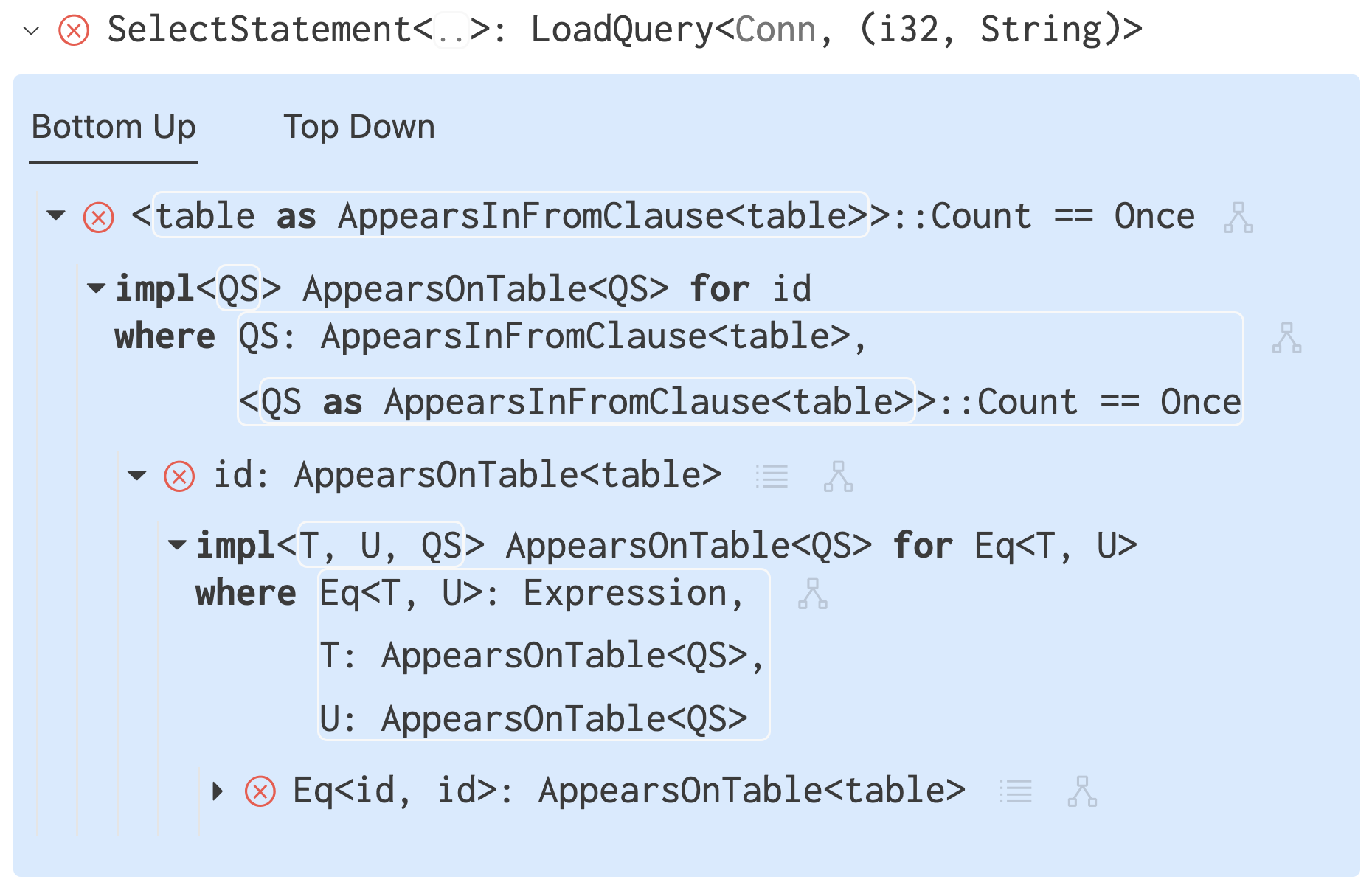}}
        \caption{Clicking an elided type to expand its definition in-place.}
        \label{fig:argus-ellipsis}
    \end{subfigure}
    \vspace{-0.5em}
    \caption{Interactions in \argus{} for expanding shortened types.}
\end{figure}

\subsubsection{\prinzipZwoo}\label{principle:2}
\textit{Instead of heuristically shortening types, show shortened types by default, but make fully-qualified types available on-demand.}

Textual diagnostics must deal with large types via a combination of pretty-printing, heuristic shortening, and file logging. \argus{} instead shortens all types by default and enables developers to contextually expand them in two ways:

\begin{itemize}
    \item Fully-qualified definition paths are removed, and only symbol names are printed by default. For example, the interface would print \rs{SelectStatement} instead of \rs{diesel::SelectStatement}. To observe the full path, the developer can hover their mouse over a symbol name, and its path appears in a mini buffer at the bottom of the page as shown in \Cref{fig:argus-minibuffer}.
    
    \item Trait parameters, impl block quantified types, and impl block where-bounds are hidden by default, represented by an ellipsis. For example, \argus{} will display \rs{SelectStatement<..>} instead of \rs{SelectStatement<FromClause<table, ...>>}. The developer can click the ellipsis to expand out the hidden content as shown in \Cref{fig:argus-ellipsis}.
\end{itemize}

\begin{figure}
    \begin{subfigure}{0.45\textwidth}
        \boxf{\includegraphics[width=\textwidth]{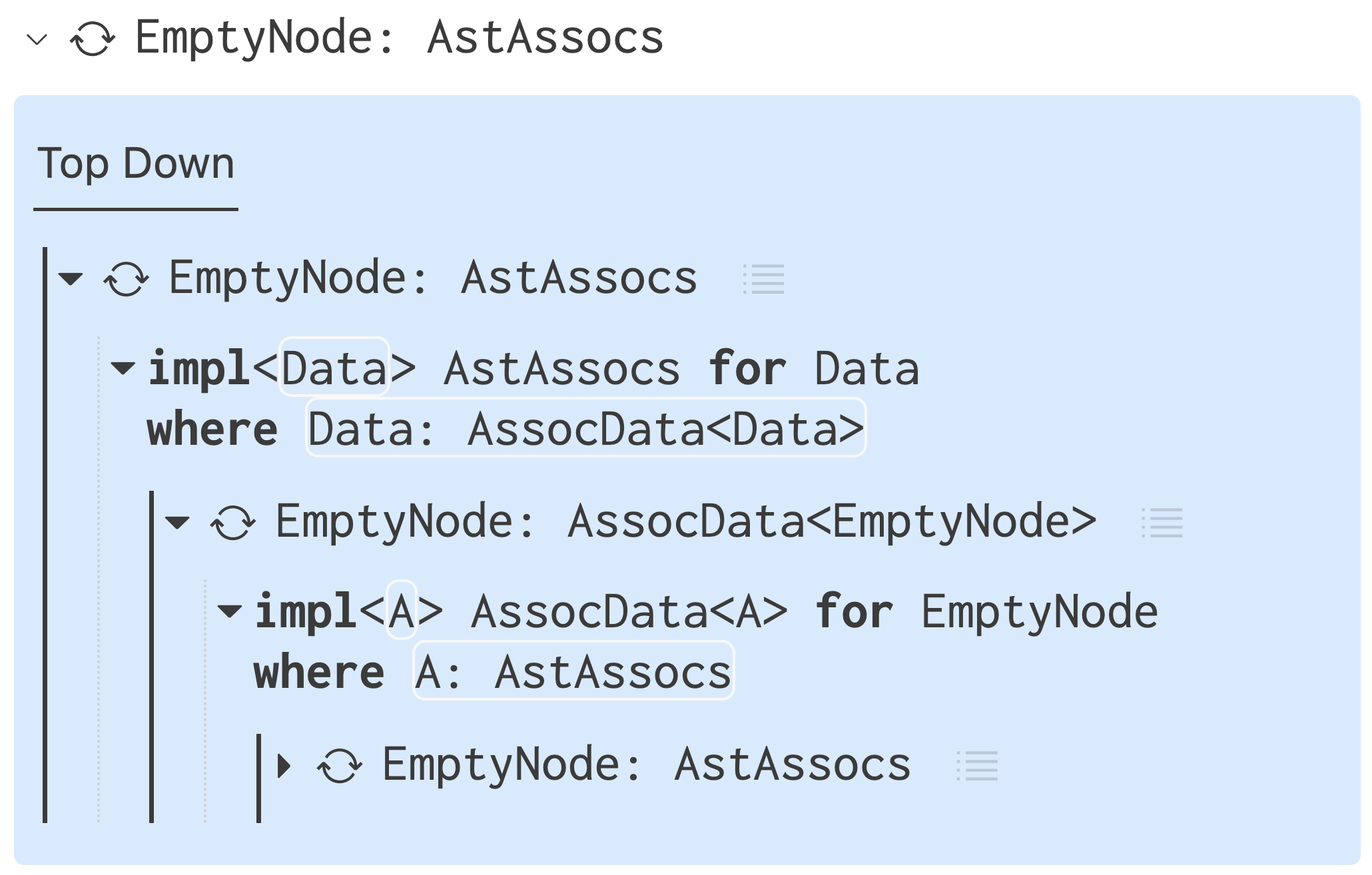}}
        \caption{Contextual information is hidden by default, showing only the core inference tree structure.}
        \label{fig:ast-overflow-argus}
    \end{subfigure}
    \hfill
    \begin{subfigure}{0.45\textwidth}
        \boxf{\includegraphics[width=\textwidth]{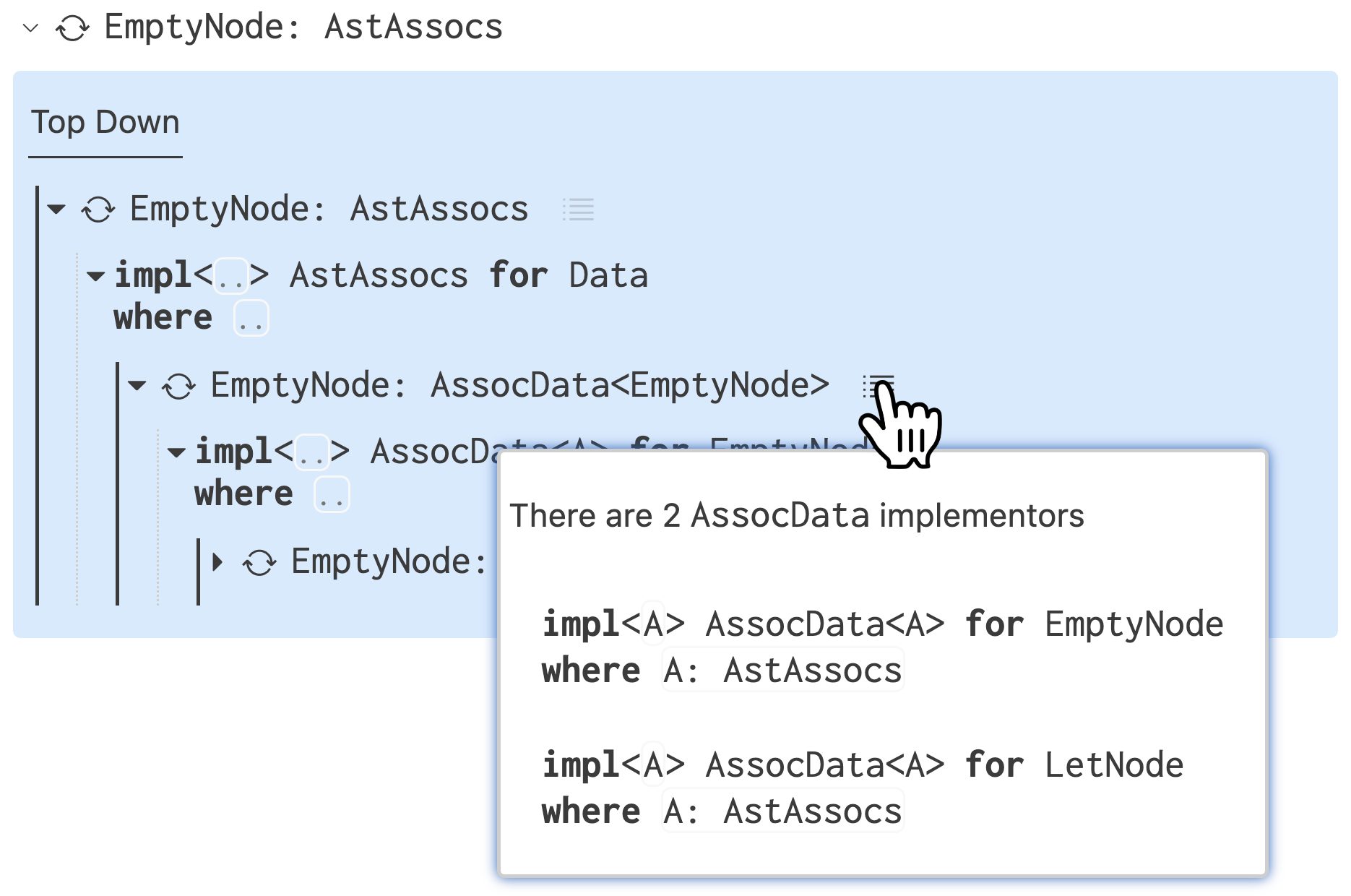}}
        \caption{The developer can query any trait bound for the full set of impl blocks for that trait, shown in a popup.}
        \label{fig:argus-impl-list}
    \end{subfigure}
    \vspace{-0.5em}
    \caption{Interactions in \argus{} for accessing contextual information about types and traits.}
\end{figure}

\subsubsection{\prinzipDruu}\label{principle:3}
\textit{Instead of interleaving the trait inference steps with auxiliary information, enable developers to access auxiliary information on-demand through contextual links.}


The core visualization of the inference tree in \argus{} just shows the information strictly needed for the inference process: trait bounds and impl blocks. As shown in \Cref{fig:ast-overflow-argus}, this declutters the inference tree so that previously-obscured relationships like the overflow from \Cref{sec:case-study-ast} become simpler to visually track.
All auxiliary data is instead accessible through hyperlinks and popup windows. Specifically:

\begin{itemize}
    \item Developers can command-click any symbol to jump to its definition in the adjacent code editor.
    \item Developers can click a button next to each trait to access the list of impl blocks for that trait, as shown in \Cref{fig:argus-impl-list}.
\end{itemize}

\begin{figure}
    \centering
    \begin{subfigure}[t]{0.49\textwidth}
        \boxf{\includegraphics[width=\textwidth]{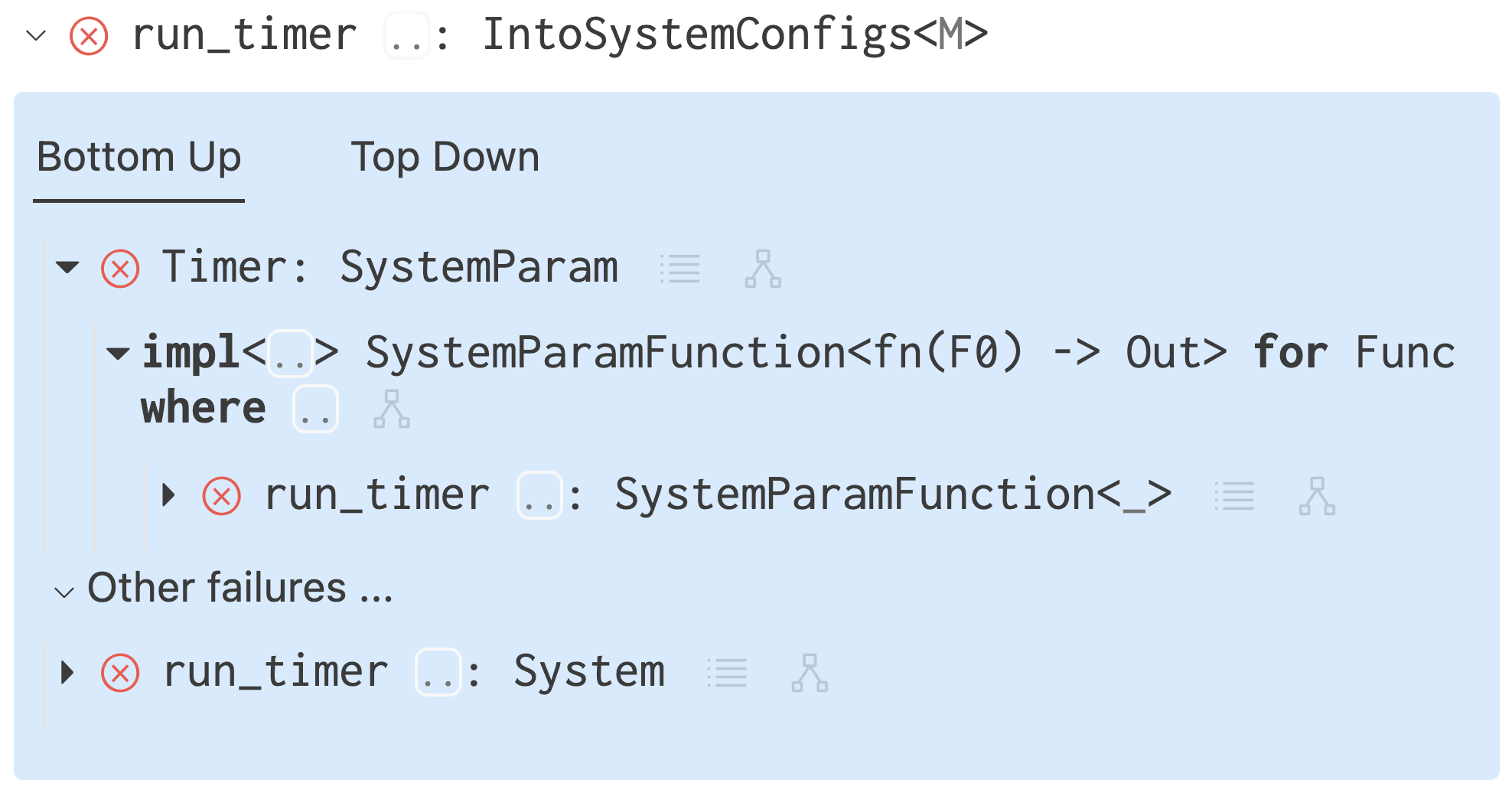}}
        \cprotect\caption{The bottom-up view shows the deepest failed predicates, and unfolds the parents.}
        \label{fig:bevy-argus-bu}
    \end{subfigure}
    \hfill
    \begin{subfigure}[t]{0.49\textwidth}
        \boxf{\includegraphics[width=\textwidth]{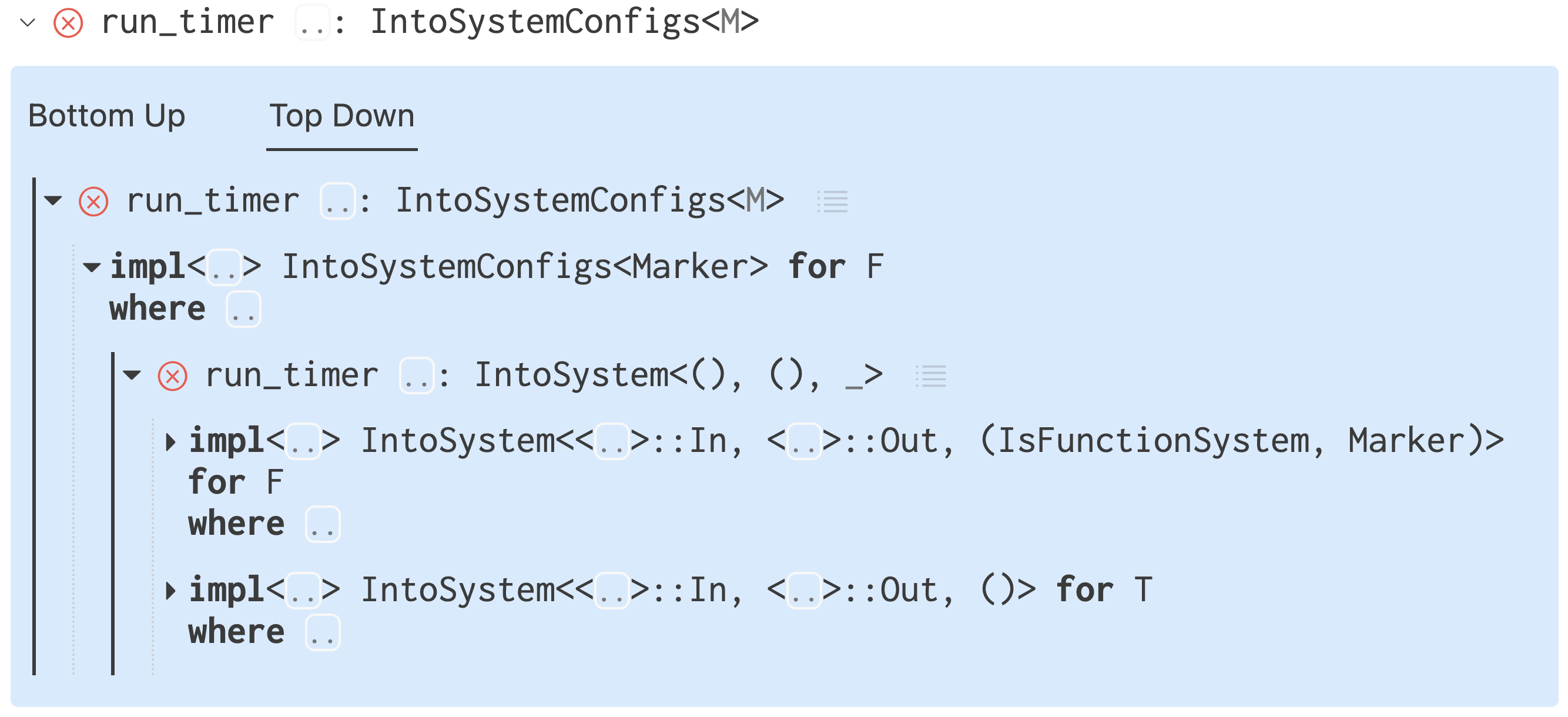}}
        \cprotect\caption{The top-down view shows the root failed predicate, and unfolds its children.}
        \label{fig:bevy-argus-td}
    \end{subfigure}
    \vspace{-0.5em}
    \caption{\argus{} visualizes two different projections of a trait inference tree: bottom-up and top-down.}
\end{figure}

\subsubsection{\prinzipVier}\label{principle:4}
\textit{Instead of omitting tree-shaped information in a trait inference, provide an interface that supports exploring the trait inference as a tree.}

A tree can be visualized in dozens of ways~\cite{schulz2011treevis}, with different trade-offs for the kinds of information which are easy and hard to find and understand. For example, one could imagine visualizing the inference tree in a pannable node-link diagram, which would more effectively convey the ``10,000 foot view'' on the inference tree. We opted specifically for a nesting-based representation because we expect that a high-level view is not particularly useful for trait debugging, since a developer most often cares about finding specific nodes in the tree. Future versions of \argus{} targeted at, e.g., helping Rust compiler developers design and debug the trait system itself might benefit more from a high-level view, but here we just focus on user-space debugging.

In particular, our hypothesis is that a developer may find useful both the top-down and bottom-up views on the inference tree, depending on their specific question. A bottom-up view, as shown in \Cref{fig:bevy-argus-bu}, emphasizes most directly the possible root causes for the error. If the developer can understand these failed trait bounds without much context, e.g., by reading \rs{Timer: SystemParam} and immediately understanding the problem, then the bottom-up view most directly facilitates fault localization. If a developer cannot understand failed trait bounds out of context, one option is to iteratively unfold the parents in the bottom-up view. Alternatively, the developer can use the top-down view to get a more ``logical'' view on the situation, as shown in \Cref{fig:bevy-argus-td}. The developer reads from the visualization: we started needing to show \rs{run_timer: IntoSystemConfigs<M>}, and that required \rs{run_timer: IntoSystem<(), (), _>}, which could be satisfied in one of two ways, and so on.

\subsection{Ranking Predicates with Inertia}\label{sec:inertia}

The bottom-up view presents the innermost failing predicates in a particular sequence, which the developer presumably reads top-to-bottom. There is no inherent order to these predicates, because in theory each predicate could be the one that the developer intended to satisfy. In practice, we believe that some predicates are on average more likely than others to be the root cause, and that likelihood can be analyzed just from the structure of the predicate.

\begin{figure}
    \centering
    \includegraphics[width=\linewidth]{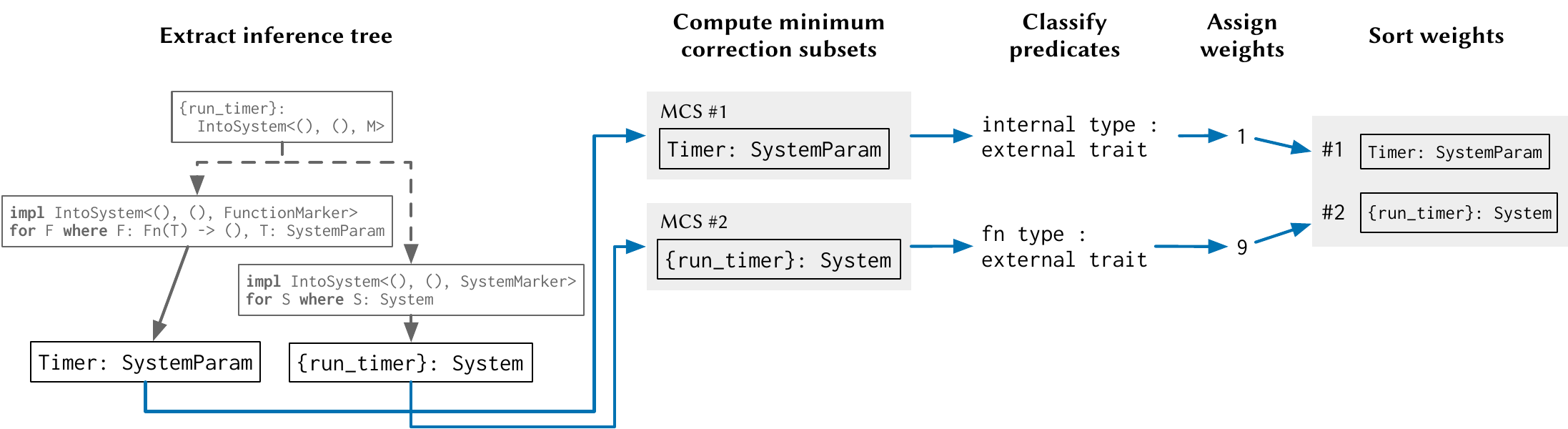}
    \caption{An example of applying the inertia heuristic to the Bevy inference tree in \Cref{fig:bevy-example-diagram}. Given an extracted inference tree, \argus computes the set of smallest subsets required to satisfy the root obligation, classifies each predicate based on its structure, assigns a human-decided weight to each category, and sorts the weights.}
    \label{fig:inertia-diagram}
\end{figure}

Our theory is that the correct fix to a failed trait error on average involves the fewest modifications to program elements, such as type definitions and trait implementations. This theory motivates our heuristic, \emph{inertia}, used by \argus{} to sort failed predicates as shown in \Cref{fig:inertia-diagram}. Inertia models the complexity of the patch required to fix a failed predicate. For Rust, we designed the inertia heuristic to reflect two common sources of complexity in fixing trait errors:
\begin{enumerate}
    \item \emph{Orphan rule:} to ensure coherence of trait implementations between libraries, Rust disallowed implementing an externally-defined trait for an externally-defined type. This means a failed trait bound requiring an external type to implement an external trait requires more changes than with local types or traits (e.g., wrapping the external type in a local newtype, or changing the external type/traits).
    \item \emph{Function traits:} higher-order functions in Rust are not generally written using function \emph{types} as in most functional languages, but rather function \emph{traits} due to the interaction of closures and ownership in Rust. Traits implemented for functions are written as blanket implementations like \rs{impl<F: Fn(A) -> B> Foo for F} as opposed to hypothetically \rs{impl Foo for fn(A) -> B}. As a result, function implementations are not rejected via unification of the head type, and they often appear as failed alternatives in trait inferences. 
\end{enumerate}

For example, consider the two innermost failed trait bounds in the bottom-up view on the Bevy program in \Cref{fig:bevy-argus-bu}: \rs{{run_timer}: System} and \rs{Timer: SystemParam}. Informally, the first bound should be higher inertia because (a) it would violate the orphan rule and (b) it involves creating a new function trait. The second bound should be lower inertia because (a) \rs{Timer} is a local type and therefore does not violate the orphan rule, and (b) no referenced types are functions.

Formally, we compute inertia by first enumerating all minimum correction subsets (\mcs). An \mcs is a set of failing predicates in the trait inference tree that, if they hold true, would cause the proof to hold true. Specifically, we treat the \AndOr tree as a propositional logic formula and normalize it into disjunctive-normal form (\dnf). For each conjunct in the \dnf formula, we apply the inertia heuristic to compute a score for each predicate in the conjuct. The conjunct's final score is the sum of its predicate scores.

To score a predicate, we categorize it into one of eight categories of predicates. Three key categories are (1) coherent non-function trait bounds, (2) orphaned non-function trait bounds, and (3) function trait bounds. For example, \rs{{run_timer}: System} is in category \#3 and \rs{Timer: SystemParam} is in category \#1. We assigned each category a numeric rank based on the expected complexity of the fix, so e.g. category 1 is lower than 2 is lower than 3. This produces the sort order shown in \Cref{fig:bevy-argus-bu}. An exhaustive list of the categories and their ranking is provided in \Cref{app:inertia}.

\section{Implementation}\label{sec:implementation}

\argus is implemented as a VS Code extension that is freely available on the VS Code Marketplace and Open VSX Registry. The Rust compiler plugin is \estimate{10,393} lines of Rust code, of which \estimate{4,216} lines (\estimate{40.6\%}) are just for serializing the Rust type system to JSON. The \argus interface is \estimate{8,470} lines of TypeScript code, of which \estimate{2,327} lines (\estimate{27\%}) are just for pretty printing the Rust type system.

Beyond type serialization, the most significant implementation detail in \argus{} is how we extract the idealized \AndOr tree representation from the trait solver. One complication is that not all predicates evaluated by the trait solver represent the ``final'' predicates that should be presented to the developer, because trait solving and type checking are interleaving processes. It is possible that the Rust compiler provides a trait predicate with unknown type variables. Solving predicates happens in a fixpoint; ambiguous predicates remain in the trait solver queue until they are proved true or false, or until inference finishes, at which point all ambiguous predicates become failures. This reality is difficult for extensions like \argus because predicates re-entered into the trait solving queue are represented as new predicates. This means that \argus sees all snapshots of a predicate's evolution, and we use an implication heuristic to remove earlier predicates.

The second complication is the process by which trait solving and type inference guide each other. A good example is trait method calls. Consider the expression \rs{my_value.to_string()}. Initially, there are two unknowns: the type of \rs{my_value}, and where the method \rs{.to_string()} comes from. Say that \rs{my_value} has type \rs{Vec<i32>} and two traits \rs{ToString} and \rs{CustomToString} provide the method \rs{.to_string()}. The type inference engine may ask the trait solver to evaluate \rs{Vec<i32>: ToString}, but this predicate is \textit{speculative.} If the predicate fails, the inference engine may ask the trait solver to evaluate \rs{Vec<i32>: CustomToString}. The issue is that all predicates, regardless of whether they are soft or hard constraints, look identical to external compiler plugins. \argus uses a heuristic to reverse-engineer the predicates evaluated in a program and attempts to show as few as possible. However, the version of \argus deployed in the user study showed potentially more failing predicates than necessary.

Finally, the grammar for \tLang contains three possible predicates (trait bounds, projections, outlives-constraints), but there are actually fourteen in the Rust compiler implementation. Several of the included predicates are important details specific to Rust, but we don't want to expose them to unsuspecting developers. \argus provides a toggle setting where developers can see the full range of predicates.

Beyond a higher \emph{quantity} of predicates, the compiler also contains \emph{stateful} predicates, such as $\texttt{NormalizesTo}~\term~\tvar$. This predicate is the Rust equivalent of Prolog unification, except that normalization is unidirectional. Within the compiler this predicate is used semantically like a function, where the expression \term is normalized, and the expression is written into the unconstrained type variable $\tvar$. From this perspective, neither is the predicate useful nor is its subtree. \argus therefore cannot treat the entire inference tree as a tree, but rather some predicates must be treated as stateful nodes whose values can be captured only after their subtrees execute.

\section{Evaluation}\label{sec:evaluation}

The central question of our evaluation is: how does \argus{} actually influence a Rust programmer's process of debugging complex trait errors? We explore this question in three parts:
\begin{itemize}
    \item RQ1: How does \argus{} affect the overall time to localize and fix a trait error?
    \item RQ2: How do the features of the \argus{} interface individually affect a programmer's debugging process?
    \item RQ3: How useful is the inertia heuristic in improving the rank order of the bottom-up view?
\end{itemize}

We answer RQ1 and RQ2 by conducting a user study of $N = 25$ Rust programmers debugging a variety of trait-related errors both with and without \argus{}. We evaluate RQ1 quantitatively by measuring time-on-task, and RQ2 qualitatively by observing themes in participants' use of the tool. We answer RQ3 by running an experiment to quantitatively compare the relative efficacy of different predicate orderings given a ground truth specification of the correct fault.

\subsection{User Study}

The goal of this study was to compare Rust programmers' trait debugging strategies both with and without \argus{} in a variety of domains on relatively self-contained tasks.

\subsubsection{Methodology}

\paragraph{Participants} We recruited participants from three main sources: a mailing list of Rust learners, the Rust subreddit, and the Rust Zulip. Each source provides Rust developers of different knowledge levels. The mailing list contains people with minimal Rust knowledge, the Rust subreddit contains a diverse range of experiences, and the Rust Zulip channel is mostly Rust experts and those working on the language itself. We recruited \estimate{11} participants for a trial study. Participant feedback was used to improve the materials and instructions for the final study. We recruited $N=\estimate{\NumParticipants}$ participants for the final study. Participants had a median \estimate{11} years of programming experience (min: \estimate{2}, max: \estimate{39}), and a median \estimate{3} years of Rust experience (min: \estimate{1}, max: \estimate{9}). 

The study design was reviewed by our university IRB and determined not to meet their definition of human subjects research. We nonetheless took reasonable precaution when designing and executing our user study. No personal identifiable information was collected outside of the participant's audio and screen share during the study session. Participants were compensated \$20.

\paragraph{Materials} We created seven debugging tasks to cover a range of domains and types of trait problems. Each task consisted of a Rust crate containing one or more trait-related type errors, such as the ones shown in \Cref{sec:case-studies}. The tasks contained an average of \estimate{62} lines of application code. We used two types of libraries:
\begin{itemize}
    \item \emph{Real libraries:} widely-used Rust libraries that make heavy use of traits, specifically: the web framework Axum~\cite{axum}, the game engine Bevy~\cite{bevy}, and the SQL query builder Diesel~\cite{diesel}. These libraries contain an average of \estimate{25,771} lines of code.
    
    \item \emph{Synthetic libraries:} bespoke libraries created by us for this experiment. \rs{brew} provides an API for creating potion recipes from various plant ingredients, with invalid recipes ruled out by trait-based rules. \rs{space} provides an API to construct intergalactic flight plans, with invalid flight plans also ruled out by traits. These APIs closely mirror the designs of Axum, Bevy, and Diesel. These libraries contain an average of \estimate{721} lines of code.
\end{itemize}

Tasks involving real libraries are maximally ecologically valid, i.e., correspond to realistic problems that Rust developers encounter. However, real libraries introduce confounds: participants may have prior experience with the library, and the quality of the documentation (e.g., prose explanation and code examples) may influence task performance. The synthetic libraries control for these factors: participants cannot have prior experience with the libraries, and the libraries only use automatically-generated documentation via Rustdoc.

For each real library, we looked at community resources and selected errors that represent common beginner mistakes. We then constructed each task by injecting a fault into a well-typed program. For example, for Bevy we used the Unofficial Bevy Cheat Book~\cite{BevyCheatBookCommonMistakes}, which contains a section titled \textit{Obscure Compiler Errors}. One entry is ``Using a resource type directly without a \rs{Res} or \rs{ResMut}.'' We then took one of the Bevy example applications and removed the \rs{ResMut} wrapper around the parameter of a system function. For the synthetic libraries, we injected faults that mirror the ones in the real libraries. The full set of study materials is provided in the supplementary materials. 

\paragraph{Procedure} Participants were asked to solve a series of four debugging tasks via Zoom over the span of one hour. Study sessions were conducted during one and a half months, beginning in September 2024 and running through mid-October. All sessions were recorded.

Before arriving at their session, participants read an \argus{} tutorial and installed the \argus{} VS Code extension. At the start of the session, we verbally confirmed with participants that they completed the preparation. We gave participants time to ask questions about traits, the tutorial, or \argus. We then gave a live demonstration of \argus using the first problem from the tutorial. In the demonstration, the study administrator dictated editor actions that were carried out by the participant; each action was accompanied by a reason. For example, ``Please hover over the symbol \rs{Handler} in the \argus interface --- I want to see in which module it is defined.''

During the study, participants were given four tasks drawn randomly from the available seven. A maximum of ten minutes was allotted per task. All debugging resources were allowed, including Google, StackOverflow, and AI chatbots / coding assistants. Participants completed four tasks total, two in each condition: with \argus, and without \argus. Task order was blocked by condition, so participants did both with-\argus{} tasks and then both without-\argus{} tasks, or vice versa based on random assignment. We asked participants to think aloud and specify (1) when they had localized the error, and (2) when they had fixed the error.

\paragraph{Analysis} For each task in each session we determined two values: time-to-fix and time-to-localize. Time-to-fix is relatively straightforward: we identified when the participant provided a solution that solved the type error consistent with the problem specification. This required some qualitative analysis to distinguish trivial fixes (e.g., deleting the ill-typed code) from true fixes. Note that both times are measured from the start of the task, so time-to-fix is always strictly greater than time-to-localize.

Although we asked participants to say when they localized an error, not all did, so we qualitatively coded a localization time for each task. We consider a participant to have localized a fault once they have identified the fault and started to work on a fix for the fault. For instance, in one of the Bevy tasks, the fault is that a type \rs{Assets<Mesh>} does not implement a trait \rs{SystemParam}. We would look for indicators that the participant identified the specific issue with \rs{Assets<Mesh>}, as opposed to the entire function or unrelated parameters. We would also look for indicators that the participant was determining how \rs{Assets<Mesh>} could change to implement \rs{SystemParam}.

To evaluate our ability to consistently code for time-to-localize, the second author independently coded for this variable in \estimate{20} randomly selected tasks from the dataset. The correlation between raters was \estimate{$r = 0.998$} with a mean absolute deviation of \estimate{$34s$}. Therefore, we believe this qualitative metric can be coded with enough objectivity to be worth analyzing.

\subsubsection{Results}\label{sec:results}

\paragraph{RQ1: How does \argus{} affect overall time to localize and fix a trait error?} 

\begin{figure}
    \centering
    \begin{subfigure}[t]{0.24\textwidth}
        \includegraphics[width=\textwidth]{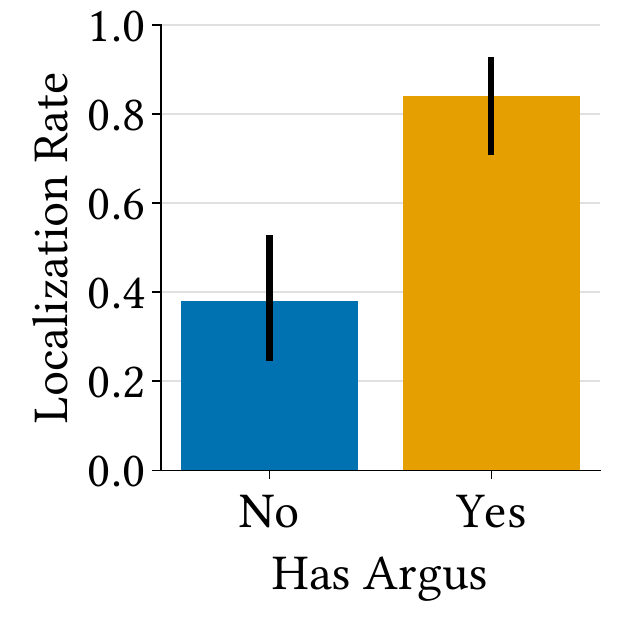}
        \caption{Localization Rate}
        \label{fig:localization-rate}
    \end{subfigure}
    \begin{subfigure}[t]{0.24\textwidth}
        \includegraphics[width=\textwidth]{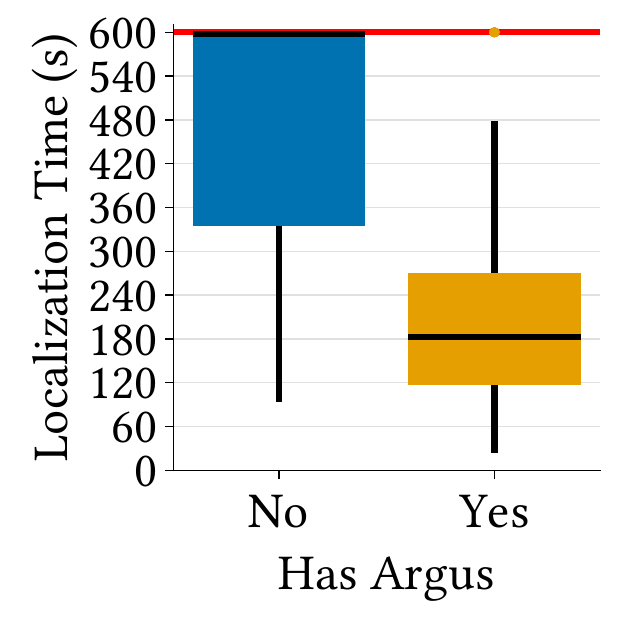}
        \caption{Localization Time}
        \label{fig:localization-duration}
    \end{subfigure}
    \begin{subfigure}[t]{0.24\textwidth}
        \includegraphics[width=\textwidth]{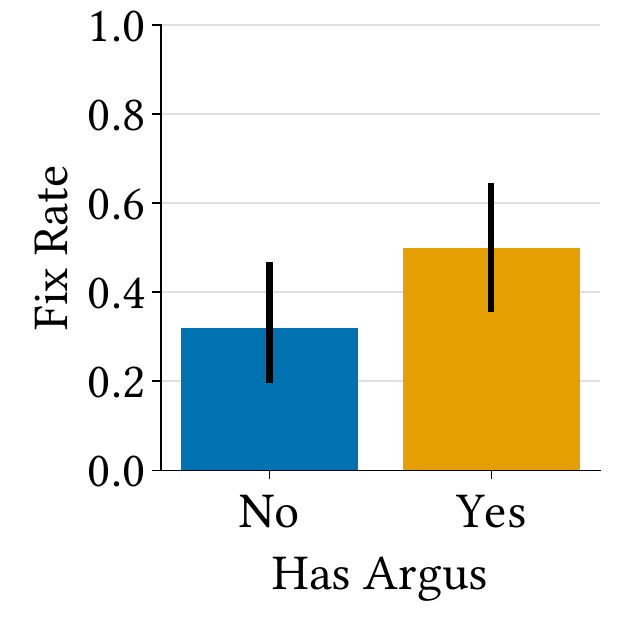}
        \caption{Fix Rate}
        \label{fig:fix-rate}
    \end{subfigure}
    \begin{subfigure}[t]{0.24\textwidth}
        \includegraphics[width=\textwidth]{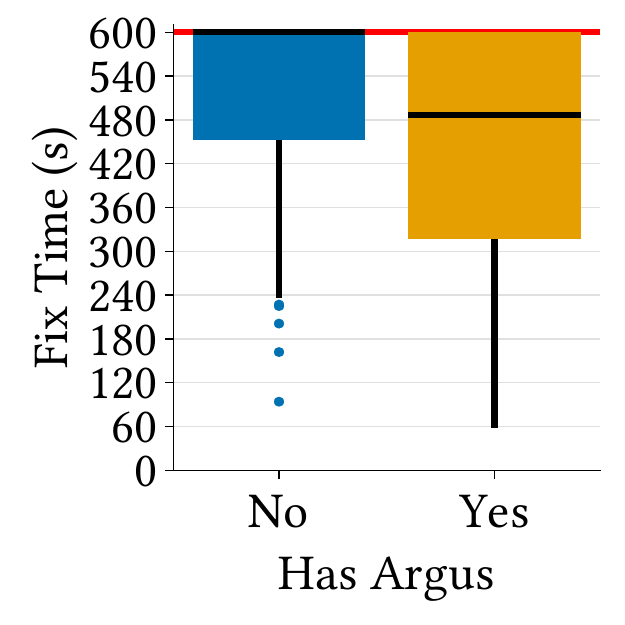}
        \caption{Fix Time}
        \label{fig:fix-duration}
    \end{subfigure}
    \caption{Distributions of localization/fix rates/times. Error bars on rates are a 95\% binomial proportion confidence interval.}
    \label{fig:eval-overview}
\end{figure}

For both metrics, we consider its overall rate (did a participant localize/fix an error), and its overall duration (when did they localize/fix the error, capped at 10min). Localization rate and time are visualized in \Cref{fig:localization-rate,fig:localization-duration}. For the localization rate, participants localized the error with \argus{} in \estimate{\WithArgusLocalize} of cases (\estimate{95\% $\ci = [\WithArgusCILoLocalize, \WithArgusCIHiLocalize]$}), and localized without in \estimate{\NoArgusLocalize} of cases (\estimate{$\ci = [\NoArgusCILoLocalize, \NoArgusCIHiLocalize]$}), a difference of \estimate{46pp} or \estimate{\LocalizedRateFactor} more cases. Using a chi-square test, this effect is statistically significant (\estimate{$\chi(1, 100) = 22.24, p < 0.001$}). 

For the localization time, participants localized the error with \argus{} in a median \estimate{3m3s} (\estimate{$\ci = [2m28s, 3m46s]$}) and without was \estimate{9m58s} (\estimate{$\ci = [7m40s, 10m]$}), a difference of \estimate{6m55s} or \estimate{$3.3\times$} faster. Using a Kruskal-Wallis test, this effect is statistically significant (\estimate{$\chi(1, 100) = 31.39, p < 0.001$}).

Fix rate and time are visualized in \Cref{fig:fix-rate,fig:fix-duration}. For the fix rate, participants fixed the error with \argus{} in \estimate{\WithArgusFix} of cases (\estimate{$\ci = [\WithArgusCILoFix, \WithArgusCIHiFix]$}), and fixed without in \estimate{\NoArgusFix} (\estimate{$\ci = [\NoArgusCILoFix, \NoArgusCIHiFix]$}) of cases, a difference of \estimate{18pp} or \estimate{$1.6\times$} more cases. Using a chi-square test, this effect is borderline statistically significant (\estimate{$\chi(1, 100) = 3.35, p = 0.07$}). To account for the within-subjects design, we further use a generalized linear model with condition as a fixed effect and participant ID as a random effect. Under this model, the effect is statistically significant (\estimate{$p = 0.03$}). 

For the fix time, participants fixed the error with \argus{} in a median \estimate{8m7s} ($\estimate{\ci = [6m30s, 10m]}$), and fixed without in \estimate{10m} (\estimate{$\ci = [9m52s, 10m]$}), a difference of \estimate{1m53s} or \estimate{$1.2\times$} faster. Using a Kruskal-Wallis test, this effect is statistically significant (\estimate{$\chi(1, 100) = 5.04, p = 0.02$}).

\paragraph{RQ2: How do the features of the \argus{} interface individually affect a programmer's debugging process?}\label{subsubsec:qualitative-analysis}

We answer RQ2 using qualitative observations from our user study, considering each design principle in turn.

\begin{enumerate}
    \setlength{\parindent}{\enumerateparindent}
    
    \item\hyperref[principle:1]{\prinzipEis}:\enskip Participants iteratively unfolded the trait inference tree to varying depths, supporting the claim that there is no one-size-fits-all depth that is optimal for diagnostics. Participants tended towards either unfolding a few steps, or unfolding the entire sequence at once. In the the former case, participants stopped early if they felt they had enough contextual information to localize an error or discard a debugging hypothesis. In the latter case, participants reported a preference for seeing all the data at once, however, they tended to spend more time debugging irrelevant information. 
    
    When first opened, the \argus interface shows the list of all bottom-up predicates together and by collapsing the inference steps the related information is viewable side by side. Not all participants started their exploration in the same place, but participants frequently collapsed inference steps after they concluded the information was no longer important. From these observations, we infer that giving participants the choice to unfold sequences is important in the exploratory phase of localization.

    \item\hyperref[principle:2]{\prinzipZwoo}:\enskip 
    For the tasks in our study, we did not observe an instance when a participant needed fully-qualified types to localize or fix a trait error, suggesting that presenting shortened types by default reduced visual noise without an adverse impact on debugging ability.

    Conversely, the compiler diagnostics contain mostly fully-qualified types. Very few participants read the full error message. Developers using VS Code's diagnostic tooltips were especially likely to skip reading the diagnostic, as most of the diagnostic overflows the tooltip and is rendered offscreen.
    
    \item\hyperref[principle:3]{\prinzipDruu}:\enskip 
    We observed that many developers preferred to look at types in in-editor source code rather than in online documentation. Participants searched source code in $\estimate{73\%}$ of tasks, while documentation was opened in only \estimate{31\%} of tasks.

    Several participants reported that if \argus was useful for nothing else, they would \textit{still} use it for access to the source hyperlinks. This observation suggests that compiler diagnostics could benefit from a rich text representation that hyperlinks types to their source definitions. 
    
    \item\hyperref[principle:4]{\prinzipVier}:\enskip  
    Most participants used both the bottom-up and top-down views, with a general preference for the bottom-up view (which is presented by default). Participants used the top-down view in \estimate{24\%} of tasks. Of those who used the top-down view, most said they preferred it for the additional context it provided.
    In the bottom-up view, participants generally explored the failed trait obligations from top to bottom, supporting the need for a sorting heuristic like inertia to optimize developer effort.
    
    The specific root causes in the inference tree provide participants with information they otherwise would not have in cases such as Bevy. For tasks involving a branching inference tree, we analyzed whether each participant identified that the root cause trait (e.g., \rs{SystemParam}) was used in the inference tree (not even that the trait was the root cause, just that it exists in the problem). Without \argus, participants only identified the trait in \estimate{29\%} of cases. Recall that the key trait is absent from the compiler's diagnostic, but useful to localize the error.
    
    Several participants reported feeling overwhelmed by the additional information surfaced by \argus. Debugging a trait error as a tree was itself a novel idea to many participants. Because \argus exposes so much information, some participants got lost in the data and ended up debugging non-issues. It is possible that these issues can be ameliorated with more instruction and further use, but we will analyze how the interface is used by the community.
\end{enumerate}

\subsection{Inertia Analysis} 

The goal of the inertia heuristic is to increase the likelihood that the root cause of a trait failure appears near the top of the bottom-up tree view. To evaluate the efficacy of the heuristic (RQ3), we compare inertia against two categories of alternatives:

\begin{itemize}
    \item \emph{Against the Rust compiler's diagnostics.} Because the compiler's diagnostics do not describe branch points, the compiler may report a failing trait bound that is higher up in the inference tree than the root cause (see \Cref{sec:case-study-bevy}). In this comparison, we ask: what is the minimal number of inference steps a developer would have to manually trace to reach the root failure?
    \item \emph{Against simpler heuristics.} We can consider the \argus{} bottom-up view but ranked using simpler heuristics than inertia. In this comparison, we ask: if sorted by a given metric, how far down from the top would a developer have to read before reaching the root failure? We specifically consider \estimate{two} heuristics: \estimate{depth of predicate in the inference tree, and number of uninstantiated inference variables in the predicate.}
\end{itemize}

Additionally, one concern with our particular inertia heuristic is the step which converts the trait inference tree into a propositional logic formula in disjunctive-normal form (\textsc{dnf}). Normalizing the tree into \textsc{dnf} is an exponential operation, which could theoretically mean significant slowdowns for larger trait inference trees. To evaluate this performance concern, we measured the normalization time on the trait inference trees in our dataset.

\subsubsection{Methodology}

For both comparisons, we need a dataset of Rust programs containing trait errors where a specific failed trait bound can be blamed as the root cause of the error. We sourced the programs from Semmler's~\cite{weiznich-grant} database of \estimate{25} Rust programs with complex trait errors. For each program, we manually identified the trait bound in its inference tree that corresponded to the root cause of the error. We removed \estimate{8} programs for a few reasons: \estimate{2} for not having a clear program intention and error cause, \estimate{2} that are well-typed but fail to compile due to bugs in the Rust compiler, \estimate{2} for not being actual trait errors, and \estimate{2} that crash the Rust compiler. Therefore, the final suite has a total of \estimate{17} programs.

For each program, we compared against the Rust compiler by generating the trait inference tree, and counting the number of nodes between the compiler's most-specific reported error and the root cause. We compared against the alternative heuristics by computing the index of the root cause in the list sorted by each heuristic. In both cases, the optimal value is 0.

To measure performance, we profiled the time spent in \dnf normalization for each program in the dataset. Performance was measured on a 2023 MacBook Pro M3 laptop with 32GB RAM.

\subsubsection{Results}

\begin{figure}
    \centering
    \begin{subfigure}[t]{0.49\textwidth}
        \includegraphics[width=\textwidth]{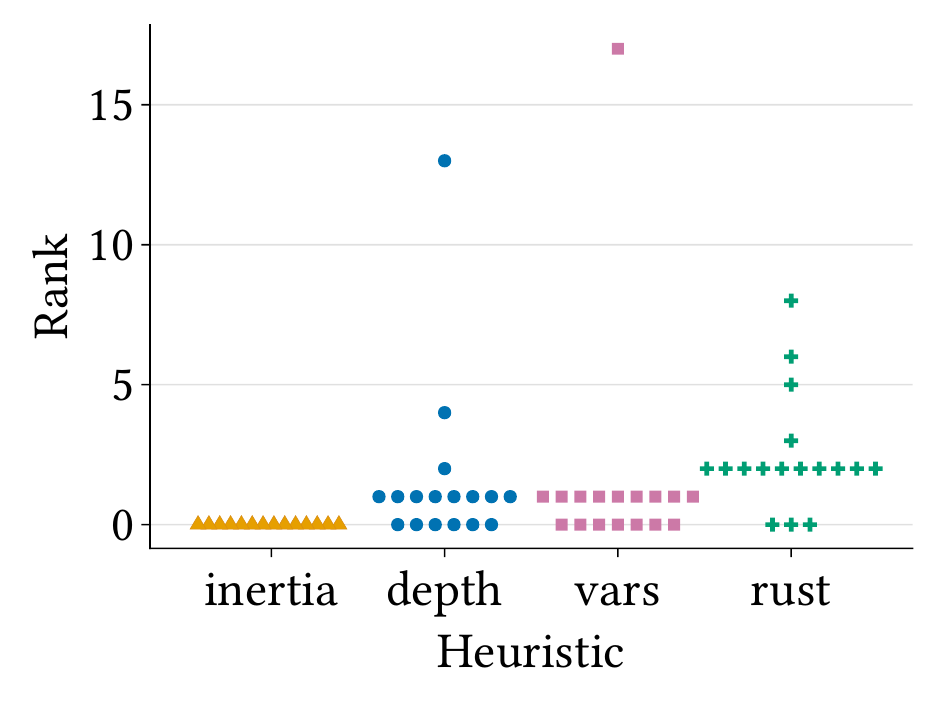}
        \caption{The distance to the root cause for the inertia heuristic, baseline heuristics, and Rust compiler diagnostic.}
        \label{fig:eval-diagnostic-precision}
    \end{subfigure}
    \hfill
    \begin{subfigure}[t]{0.49\textwidth}
        \includegraphics[width=\textwidth]{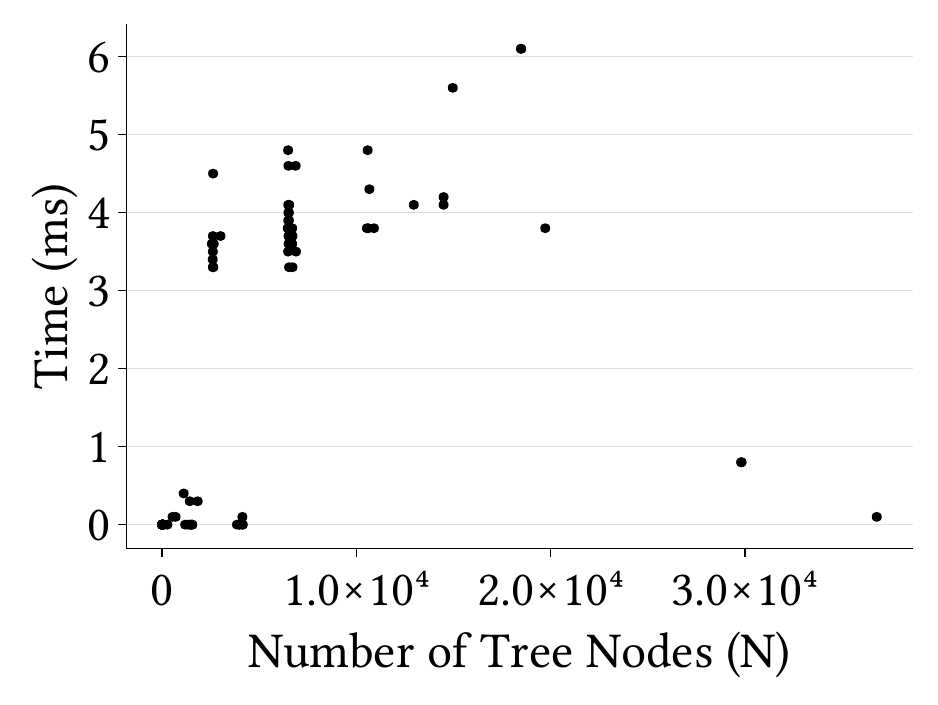}
        \caption{Time spent normalizing the trait inference trees from the test suite into \dnf.}
        \label{fig:eval-dnf-times}
    \end{subfigure}
    \caption{Inertia experiment results.}
\end{figure}

\Cref{fig:eval-diagnostic-precision} shows the distribution of distances for each approach. The median distance for each approach is \estimate{0} for inertia, \estimate{1} for predicate depth, \estimate{1} for number of inference variables, and \estimate{2} for the Rust compiler diagnostic. That is, our inertia heuristic accurately sorts the root cause to the top for every case in this particular dataset, while the other heuristics make mostly small and sometimes significant errors.

\Cref{fig:eval-dnf-times} shows the distribution of normalization time plotted against size of inference tree. The trees in our evaluation have a median size of $2,554$ nodes (min=$1$, max=$36,794$), and take a median $0.1ms$ (min<$0.001ms$, max=$6.1ms$) to normalize.

\subsection{Threats to Validity}

\subsubsection{Internal Validity}
Standard measures were taken to account for threats to internal validity. To account for sequencing effects, we randomized both the order of tasks the order of conditions (with vs. without \argus{}). Participants received training in \argus{} before completing any tasks, ensuring that participants did not receive different levels of exposure to trait debugging based on whether they used \argus{} first or last.

\subsubsection{External Validity}
We designed the tasks used to evaluate \argus{}, which could introduce bias by picking tasks favorable to our system. We combated this bias by grounding our task selection in problems identified by other people, not inventing totally new kinds of trait-related problems. Additionally, we recruited from a broad pool of Rust developers (not just, e.g., university students) so our results more likely reflect the effect of \argus{} in the general population of potential users.

Additionally, we designed the evaluation tasks to be comparable to our motivating examples described in~\Cref{sec:case-studies}, which could indicate that our evaluation does not generalize. We combat this by making \argus a general mechanism that can visualize any trait inference tree extracted from the compiler. We evaluate the system on at least the tasks it was designed for, but we cannot discount the possibility of the interface performing poorly on obscure trait errors found in the wild. We will analyze how the community uses the tool and conduct additional research should further hard-to-debug trait errors emerge.

\subsubsection{Construct Validity}
Localization time is a qualitatively-defined construct, which would be problematic if different people defined the point of localization differently. We checked for this consistency by measuring inter-rater reliability, finding that at least among ourselves we could consistently agree on the specific point plus/minus 30 seconds.

\section{Related Work}\label{sec:related-work}

\argus{} follows in a long line of systems designed to improve compiler diagnostics for type inference. Most prior work has focused on explaining failures in Hindley-Milner type inference, starting in the 1980s with the seminal work of \citet{wand1986findingte} on provenance-tracking for HM. Later work focused primarily on automatic fault localization by algorithmically deducing a single constraint to blame. Methods varied from SMT solving~\cite{pavlinovic2014type,loncaric2016type} to Bayesian analysis~\cite{meyers2015class} to machine learning~\cite{seidel2017learn}.

The systems more closely related to  \argus{} focus on human-centered methods of fault localization. These fall into three categories:
\begin{enumerate}
    \setlength{\parindent}{\enumerateparindent}
    \item \textit{Improved diagnostics:} These systems retain the static text representation of compiler diagnostics, but attempt to improve the diagnostics in some manner. One approach is to include provenance information, such as the OCaml flow-based diagnostics of \citet{bhanuka2023flow}. We similarly believe that communicating provenance is important, but for reasons discussed in \Cref{sec:case-studies}, static text is not the ideal medium to do so.
    
    Another approach is to include domain-specific information provided by library-level annotations, such as in the Helium subset of Haskell~\cite{heeren2003scripting}, which can improve the argumentative structure of a diagnostic\,\cite{barik2018error}. This is the current strategy being pursued by the Rust language developers, who recently added an \rs{#[on_unimplemented]} attribute for custom error messages~\cite{diagnostic-rfc}. For example, the Bevy diagnostic in \Cref{fig:bevy-example-diagnostic} started with the phrase \rs{`fn(Timer) {run_timer}` does not describe a valid system configuration} due to a library-level annotation in Bevy. This approach is largely orthogonal to \argus{}, which focuses on visualizing the formal structure of an inference. These approaches could also work together, e.g., if \argus{} used domain-specific messages to augment nodes in the inference tree. But we also believe that domain-specific annotations are not \emph{sufficient} for diagnosing trait errors in all cases, as also shown by the Bevy example (\Cref{sec:case-study-bevy}).
    
    \item \textit{Algorithmic debugging:} These systems present the developer a sequence of questions about predicates they expect to hold or not~\cite{shapiroaadebug}. This approach is conceptually similar to using the top-down view in \argus{} and iteratively unfolding nodes, entering sub-trees if one believes a given predicate should hold. \argus{} improves on algorithmic debugging both by providing alternative views on the inference tree (bottom-up), but also because a graphical interface simplifies certain interactions compared to the CLI such as backtracking and exploring alternative paths.
    \item \textit{Graphical interfaces:} These systems provide an interactive view onto the type inference process. In our review, we only identified two such published systems.  First, MrSpidey~\cite{flanagan1996spidey} is a visualizer for a set-based static analysis of Scheme. MrSpidey contextually visualizes inferred abstract values, and it explains the provenance of individual constraints by overlaying arrows onto the source code. Instead, \argus{} opts to provide a profiler-style separate visualization of the inference tree, which we believe is more useful in practice for debugging trait inferences.
    
    Second, the Chameleon IDE~\cite{fu2023chameleon} is a graphical enhancement of the Chameleon system~\cite{stuckey2003chameleon} for Haskell. Chameleon IDE is a single-step debugger, offering an interface for stepping through the effects of each constraint on a type inference problem. \argus{} does not try to present a stateful view on the inference process, but rather a projection of the final inference tree.
\end{enumerate}

The vast majority of research published in this area has no human-centered evaluation of their techniques. In our review, the only systems with user studies that report task performance are Chameleon IDE~\cite{fu2023chameleon} and OCaml flow diagnostics~\cite{bhanuka2023flow}. Notably, these studies found their tools either had extremely small effects or no significant effects on task performance, respectively. By contrast, we show in \Cref{sec:evaluation} that \argus{} significantly improves both the rates and duration for both localization and fixes.

In another sense, \argus{} is more properly placed in the dormant line of work on debuggers for logic programming. Back in the 1980s and 90s, researchers developed several tools to facilitate debugging of Prolog programs, in particular by visualizing \AndOr trees~\cite{dewar1986prolog,eisenstadt1988transparent,senay1991logic,ducasse1998opium}. \argus{} is essentially modernizing these interfaces and specializing their design to the use case of trait error debugging, while avoiding complexities of Prolog execution such as backtracking with cuts. However, if programmers continue to write ever more complex Turing-complete programs with type classes, then that line of work may provide useful ideas for visualizing type class inference as a stateful process rather than a pure natural deduction.

\section{Discussion}\label{sec:discussion}

Many prior systems have attempted to improve type inference diagnostics through increasingly sophisticated algorithms and heuristics. In this paper, we have argued that the interface matters, too. It's useful to think about inference trees as data and inference diagnostics as data visualization, especially for type classes. The results back up this argument --- \argus{} saves developers time and energy they would otherwise waste scrounging around the documentation. In part, this is because \argus{} empowers developers to grapple directly with the logical structure of a trait inference, rather than falling back on heuristic reasoning based on similarity to examples. Looking forward, we hope to extend \argus{} to other sub-tasks of debugging (\Cref{sec:beyond-localization}) and to other languages (\Cref{sec:beyond-rust}).

\subsection{Trait Debugging Beyond Localization}\label{sec:beyond-localization}
\argus primarily facilitates localization, or identifying the root cause of a trait error. However, localization is only part of debugging. As illustrated in \Cref{fig:eval-overview}, many participants in our study could use \argus to successfully localize an error, but still fail to fix the error. Future iterations of \argus would ideally present more information that also facilitates fixes.

One such feature already in \argus is the ability to query for the implementers of a trait, as shown in \Cref{fig:argus-impl-list}. For example, if a user localizes a failed predicate like \rs{Timer: SystemParam} in the Bevy example, then the user can inspect the implementers of \rs{SystemParam} to potentially find alternatives like the \rs{ResMut<...>} type. While better than nothing, this strategy is still limited because additional context is likely needed to select the appropriate implementation. Bevy provides about 30 other implementations of \rs{SystemParam}, and it requires understanding of the library design and application design to pick among them.
More generally, an open question is how to provide domain-specific feedback by making an educated guess at the user's intention from context. The \rs{#[on_unimplemented]} feature discussed in \Cref{sec:related-work} is one approach, but it cannot capture the full range of effective diagnostics as we have discussed.

\subsection{Trait Debugging Beyond Rust}\label{sec:beyond-rust}

Rust's traits represent a particular configuration in the broader design space of type classes, and the Rust compiler uses a particular diagnostic approach. This raises the question: are the problems in \Cref{sec:case-studies} caused by Rust's specific design, or are they more fundamental to type classes? To what extent would the ideas in \argus be useful in other languages? We expect that for cases like the Diesel example (\Cref{sec:case-study-diesel}) and the AST example (\Cref{sec:case-study-ast}) that the problems described are fairly general. Every type class system involves chains of inferences, and every textual diagnostic must somehow format that inference chain along with auxiliary information like source-mapping.

The Bevy example (\Cref{sec:case-study-bevy}) is more interesting because the core concept can be encoded differently into different type class designs. Recall that the key problem with the Bevy example in Rust is the use of an inferred marker type to distinguish otherwise-overlapping trait implementations. We examined how this problem changes when implemented in Scala, Lean, and Haskell:
\begin{itemize}
    \item Scala's implicits seem most similar of the three to Rust's traits. Scala similarly requires a marker type to avoid conflicting \rs{given} blocks. When confronted with a branch point, Scala's diagnostics curiously seem to pick a single branch and assume the developer intended that branch, but we could not determine Scala's algorithm for selecting the best branch.
    \item Lean's type classes are more expressive than Rust, as Lean has both first-class variables and permits overlapping instances (disambiguated via either names or via a numeric priority system). If we use the approach of encoding the marker type as a metavariable, then Lean returns a pithy diagnostic that again halts at the branch point: \rs{failed to synthesize IntoSystem (Timer -> Unit) ?m.619} with no further explanation.
    \item Haskell's type system offers more mechanisms for manipulating type class inference than Rust, Scala, or Lean. One way to model the Bevy API in Haskell is to represent the marker type as a type family. 
    When encoded this way, the type family eliminates the branch point in the trait inference tree, because the type of the marker is \emph{computed} via the type family as opposed to \emph{inferred} from constraints. Subsequently, the Haskell encoding of the Bevy program generates a diagnostic that is comparably specific to \argus{}, i.e., that \rs{Timer: SystemParam} is the root cause.

\end{itemize}

In sum, the design of a language's type class system will certainly affect the level of quality achievable in the language's diagnostics. More generally, modern languages increasingly use some form of search during compilation, whether that's automated theorem proving at the type-level (e.g., type classes, refinement types, tactic-based proofs) or program search/synthesis at the expression-level (e.g., supercompilation, polyhedral analysis, e-graph optimization). From the developer's perspective, all of these techniques share a common thread: it's great when they work, and hard to understand when they don't. Language designers need to understand the usability trade-offs inherent to search-based methods: they place a greater burden on the compiler to explain its work. We hope that tools like \argus{} help broaden our community's conception of the ways a compiler can explain itself.

\section*{Data-Availability Statement}
The tutorial, study questions, raw data, and data analyses are all available in our Zenodo artifact~\cite{zenodo-artifact}. \argus is open-source software available on GitHub~\cite{argus-source}, and the IDE extension is published and freely available on the \href{https://marketplace.visualstudio.com/items?itemName=gavinleroy.argus}{VS Code Marketplace} and \href{https://open-vsx.org/extension/gavinleroy/argus}{Open VSX Registry}.

\begin{acks}
    This work was partially supported by the DARPA under Agreement No.\@ HR00112420354 and partially supported by the NSF under Award No.\@ CCF-2227863. Any opinions, findings, and conclusions or recommendations expressed in this material are those of the authors and do not reflect the views of our funders. We thank our reviewers and shepherd for their insightful and thorough feedback. Library maintainers Georg Semmler (Diesel) and Alice Ryhl (Bevy) provided early feedback and expert opinions on trait diagnostics. Members of the Rust Types Team, especially lcnr and Michael Goulet, answered numerous questions about working with the compiler and the new trait solver's interface. We are extremely grateful for the Rust community and their consistent enthusiasm to participate in user studies.
\end{acks}

\bibliography{bibs/bibfile,bibs/from_thesis}

\ifcameraready
\makeatletter
\par\bigskip\noindent\small\normalfont\@received\par
\makeatother
\fi

\newpage
\appendix
\section{Appendix}

\subsection{Inertia}
\label{app:inertia}

The exact Rust code modeling the different change types and weights is below.

\begin{lstlisting}
enum GoalKind {
  Trait { _self: Location, _trait: Location },
  TyChange,
  FnToTrait { _trait: Location, arity: usize },
  TyAsCallable { arity: usize },
  DeleteFnParams { delta: usize },
  AddFnParams { delta: usize },
  IncorrectParams { arity: usize },
  Misc,
}
\end{lstlisting}

\begin{lstlisting}
impl GoalKind {
  fn weight(&self) -> usize {
    use GoalKind as GK;
    use Location::{External as E, Local as L};
    match self {
      GK::Trait {
        _self: L,
        _trait: L,
      } => 0,

      GK::Trait {
        _self: L,
        _trait: E,
      }
      | GK::Trait {
        _self: E,
        _trait: L,
      }
      | GK::FnToTrait { _trait: L, .. } => 1,

      GK::Trait {
        _self: E,
        _trait: E,
      } => 2,

      GK::TyChange => 4,
      GK::IncorrectParams { arity: delta }
      | GK::AddFnParams { delta }
      | GK::DeleteFnParams { delta } => 5 * delta,
      GK::FnToTrait { _trait: E, arity }
      | GK::TyAsCallable { arity } => 4 + 5 * arity,
      GK::Misc => 50,
    }
  }
}
\end{lstlisting}

\ifcameraready
\setcounter{TotPages}{21}
\fi

\end{document}